\documentclass[superscriptaddress,twocolumn]{revtex4}
\usepackage{amsmath,amssymb}
\usepackage{graphicx}
\usepackage{color}
\usepackage[utf8]{inputenc}
\newcommand{\Mainz}[1]
{\affiliation{Institut f\"ur Kernphysik, University of Mainz, D-55099 Mainz,Germany}}
\newcommand{\Bonn}[1]
{\affiliation{Helmholtz-Institut f\"ur Strahlen- und Kernphysik, University of Bonn,
 D-53115 Bonn, Germany}}
\newcommand{\Regina}[1]
{\affiliation{University of Regina, Regina, Saskatchewan S4S 0A2, Canada}}
\newcommand{\Glasgow}[1]
{\affiliation{SUPA School of Physics and Astronomy, University of Glasgow,
 Glasgow G12 8QQ, United Kingdom}}
\newcommand{\Kent}[1]
{\affiliation{Kent State University, Kent, Ohio 44242-0001, USA}}
\newcommand{\Giessen}[1]
{\affiliation{II Physikalisches Institut, University of Giessen, D-3539 Giessen, Germany}}
\newcommand{\Dubna}[1]
{\affiliation{Joint Institute for Nuclear Research, 141980 Dubna, Russia}}
\newcommand{\Pavia}[1]
{\affiliation{INFN Sezione di Pavia, I-27100 Pavia, Italy}}
\newcommand{\GWU}[1]
{\affiliation{The George Washington University, Washington, DC 20052-0001, USA}}
\newcommand{\LPI}[1]
{\affiliation{Lebedev Physical Institute, 119991 Moscow, Russia}}
\newcommand{\Dalhousie}[1]
{\affiliation{Dalhousie University, Halifax, Nova Scotia B3H 4R2, Canada}}
\newcommand{\Halifax}[1]
{\affiliation{Saint Mary’s University, Halifax, Nova Scotia B3H 3C3, Canada}}
\newcommand{\UniPavia}[1]
{\affiliation{Dipartimento di Fisica, Universit\`a di Pavia, I-27100 Pavia, Italy}}
\newcommand{\Basel}[1]
{\affiliation{Institut f\"ur Physik, University of Basel, CH-4056 Basel, Switzerland}}
\newcommand{\Tomsk}[1]
{\affiliation{Laboratory of Mathematical Physics, Tomsk Polytechnic University, 634034
Tomsk, Russia}}
\newcommand{\Edinburgh}[1]
{\affiliation{School of Physics, University of Edinburgh, Edinburgh EH9 3JZ,
 United Kingdom}}
\newcommand{\INR}[1]
{\affiliation{Institute for Nuclear Research, 125047 Moscow, Russia}}
\newcommand{\Sackville}[1]
{\affiliation{Mount Allison University, Sackville, New Brunswick E4L 1E6, Canada}}
\newcommand{\Zagreb}[1]
{\affiliation{Rudjer Boskovic Institute, HR-10000 Zagreb, Croatia}}
\newcommand{\Amherst}[1]
{\affiliation{University of Massachusetts, Amherst, Massachusetts 01003, USA}}
\newcommand{\UCLA}[1]
{\affiliation{University of California Los Angeles, Los Angeles, California 90095-1547, USA}}
\newcommand{\Jerusalem}[1]
{\affiliation{Racah Institute of Physics, Hebrew University of Jerusalem, Jerusalem 91904, Israel}}
\begin{document}
\title{Experimental study of the $\gamma p\to \pi^0\eta p$ reaction with the A2 setup at the Mainz Microtron}

\author{V.~Sokhoyan}\thanks{corresponding author, e-mail: sokhoyan@uni-mainz.de} \Mainz \\
\author{S.~Prakhov}\Mainz \\ \UCLA \\
\author{A.~Fix}\Tomsk \\
\author{S.~Abt}\Basel \\
\author{P.~Achenbach}\Mainz \\
\author{P.~Adlarson}\Mainz \\
\author{F.~Afzal}\Bonn \\
\author{P.~Aguar-Bartolom\'e}\Mainz \\
\author{Z.~Ahmed}\Regina \\
\author{J.~Ahrens}\Mainz \\
\author{J.~R.~M.~Annand}\Glasgow \\
\author{H.~J.~Arends}\Mainz \\
\author{K.~Bantawa}\Kent \\
\author{M.~Bashkanov}\Edinburgh \\
\author{R.~Beck}\Bonn \\
\author{M.~Biroth}\Mainz \\
\author{N.~S.~Borisov}\Dubna \\
\author{A.~Braghieri}\Pavia \\
\author{W.~J.~Briscoe}\GWU \\
\author{S.~Cherepnya}\LPI \\
\author{F.~Cividini}\Mainz \\
\author{C.~Collicott}\Dalhousie \\ \Halifax \\
\author{S.~Costanza}\Pavia \\ \UniPavia \\
\author{A.~Denig}\Mainz \\
\author{M.~Dieterle}\Basel \\
\author{E.~J.~Downie}\GWU \\
\author{P.~Drexler}\Mainz \\
\author{M.~I.~Ferretti Bondy}\Mainz \\
\author{L.~V.~Fil'kov}\LPI \\
\author{S.~Gardner}\Glasgow \\
\author{S.~Garni}\Basel \\
\author{D.~I.~Glazier}\Glasgow \\ \Edinburgh \\
\author{I.~Gorodnov}\Dubna \\
\author{W.~Gradl}\Mainz \\
\author{M.~G\"unther}\Basel \\
\author{G.~M.~Gurevich}\INR \\
\author{C.~B.~Hamill}\Edinburgh \\
\author{L.~Heijkenskj\"old}\Mainz \\
\author{D.~Hornidge}\Sackville \\
\author{G.~M.~Huber}\Regina \\
\author{A.~K\"aser}\Basel\\
\author{V.~L.~Kashevarov}\Mainz \\ \Dubna \\
\author{S.~Kay}\Edinburgh \\
\author{I.~Keshelashvili}\Basel\\
\author{R.~Kondratiev}\INR \\
\author{M.~Korolija}\Zagreb \\
\author{B.~Krusche}\Basel \\
\author{A.~Lazarev}\Dubna \\
\author{V.~Lisin}\INR \\
\author{K.~Livingston}\Glasgow \\
\author{S.~Lutterer}\Basel \\
\author{I.~J.~D.~MacGregor}\Glasgow \\
\author{D.~M.~Manley}\Kent \\
\author{P.~P.~Martel}\Mainz \\ \Sackville \\
\author{J.~C.~McGeorge}\Glasgow \\
\author{D.~G.~Middleton}\Mainz \\ \Sackville \\
\author{R.~Miskimen}\Amherst \\
\author{E.~Mornacchi}\Mainz \\
\author{A.~Mushkarenkov}\Pavia \\ \Amherst \\
\author{A.~Neganov}\Dubna \\
\author{A.~Neiser}\Mainz \\
\author{M.~Oberle}\Basel \\
\author{M.~Ostrick}\Mainz \\
\author{P.~B.~Otte}\Mainz \\
\author{D.~Paudyal}\Regina \\
\author{P.~Pedroni}\Pavia \\
\author{A.~Polonski}\INR \\
\author{G.~Ron}\Jerusalem \\
\author{T.~Rostomyan}\Basel \\
\author{A.~Sarty}\Halifax \\
\author{C.~Sfienti}\Mainz \\
\author{K.~Spieker}\Bonn \\
\author{O.~Steffen}\Mainz \\
\author{I.~I.~Strakovsky}\GWU \\
\author{B.~Strandberg}\Glasgow \\
\author{Th.~Strub}\Basel \\
\author{I.~Supek}\Zagreb \\
\author{A.~Thiel}\Bonn \\
\author{M.~Thiel}\Mainz \\
\author{A.~Thomas}\Mainz \\
\author{M.~Unverzagt}\Mainz \\
\author{Yu.~A.~Usov}\Dubna \\
\author{S.~Wagner}\Mainz \\
\author{N.~K.~Walford}\Basel \\
\author{D.~P.~Watts}\Edinburgh \\
\author{D.~Werthm\"uller}\Glasgow \\ \Basel \\
\author{J.~Wettig}\Mainz \\
\author{L.~Witthauer}\Basel \\
\author{M.~Wolfes}\Mainz \\
\author{L.~A.~Zana}\Edinburgh \\

\collaboration{A2 Collaboration at MAMI}

\date{\today}

\begin{abstract}
 The data available from the A2 Collaboration at MAMI were analyzed to select
 the $\gamma p\to \pi^0\eta p$ reaction on an event-by-event basis, which allows
 for partial-wave analyses of three-body final states to obtain more reliable results,
 compared to fits to measured distributions.
 These data provide the world’s best statistical accuracy in the energy range from
 threshold to $E_{\gamma}=1.45$~GeV, allowing
 a finer energy binning in the measurement of all observables needed for
 understanding the reaction dynamics.
 The results obtained for the measured observables are compared to existing
 models, and the impact from the new data is checked by the fit with
 the revised Mainz model. 
\end{abstract}

\maketitle

\section{Introduction}

 The unique extraction of partial-wave scattering amplitudes and universal baryon-resonance
 parameters from experimental data, as well as their precise interpretation
 in QCD, ranks among the most challenging tasks in modern hadron physics.
 During the last decades, an enormous effort  to study baryon resonances in photoinduced
 meson production at  various laboratories has started. Very significant progress was made
 in single- and double-meson photoproduction~\cite{PDG, Anisovich2016,BeckThoma2017,CredeRoberts2013,Tiator2011,KlemptRichard2010}. 
 The important advantage of studying multimeson final
 states is the possibility of accessing cascading decays of higher-lying resonances through
 intermediate excited states, whereas single-meson production is limited to
 decays into a meson and a ground-state nucleon. In addition, multi-meson final states
 can be used for investigating the decay modes of already established resonances as well as
 for investigating the long-standing problem of ``missing resonances'', which could couple to
 intermediate states involving excited nucleons rather than to states consisting of a meson
 and a ground-state nucleon.

 Among double-meson final states, the photoproduction of two pions, in particular of $\pi^{0}\pi^{0}$
pairs, was studied extensively within the last two decades 
(see, e.g., Refs.~\cite{Dieterle2015,Sokhoyan2015EPJA,Sokhoyan2015PLB,Thiel2015,Oberle2013,Zehr2012,Kashevarov2012,Thoma2008,Sarantsev2008,Assafiri2003}). 
In comparison to the widely studied $\gamma N\to 2\pi^0 N$ reaction, the $\gamma N\to \pi^0\eta N$ reaction provides a more selective
 identification of contributing resonances and their decay modes. In particular, for the incoming-photon
 energy range from the production threshold up to $E_{\gamma} = 1.5$~GeV, various analyses indicate
 the dominance of the $D_{33}$ partial wave~\cite{Doering2006,Mainz_2008,Mainz_2010,Kashev_pi0eta_prod,Kashev_pi0eta_asym,CBELSA_pi0etap_2008,CBELSA_pi0etap_2010,CBELSA_pi0etap_2014},
 which couples strongly to the $\Delta(1700)3/2^-$ resonance close to the production threshold
 and to the $\Delta(1940)3/2^-$ at higher energies.
Furthermore, the $\eta$ meson, acting as an isospin filter, is only emitted in transitions between
either two $N^*$ or two $\Delta$ resonances, introducing additional selectivity into the investigated
decay modes. Thus, the $\gamma N\to \pi^0\eta N$ reaction is well suited for studying production of
 the $\Delta(1700)3/2^-$ resonance not only on free protons, but also on nucleons bound in nuclei,
 where the interpretation of ambiguities in resonance contributions and decays is,
 however, more complicated, compared to the free-proton case.

 So far, the unpolarized total and differential cross sections for $\pi^{0}\eta$ photoproduction on
 the free proton, as well as polarization observables with circularly and linearly polarized photon beams,
 were measured with the CBELSA/TAPS experimental setup at the ELSA
 accelerator~\cite{CBELSA_pi0etap_2008,CBELSA_pi0etap_2010,CBELSA_pi0etap_2014},
 with the A2 setup at the MAMI accelerator~\cite{Kashev_pi0eta_prod,Kashev_pi0eta_asym,Mainz_2010},
 with the GRAAL detector at the ESRF accelerator~\cite{GRAAL_2008}, and at the LNS accelerator~\cite{LNS2006}.
 The unpolarized cross section and the beam-helicity asymmetry for photoproduction of $\pi^{0}\eta$ pairs
 on the deuteron and on helium nuclei have been recently published by the A2 collaboration~\cite{Kaeser2016}.
 Further data sets, using circularly polarized photons and heavier nuclear targets (carbon, aluminum, and lead),
 were recently acquired with the A2 setup at MAMI and will be published in a forthcoming paper.

Reliable experimental measurement and theoretical analysis of the $\gamma N\to \pi^0\eta N$ reaction,
with the three-body final state, is quite challenging because of its five-dimensional phase space.
The most efficient way of analyzing such a reaction would be a partial-wave analysis (PWA)
that enables fitting experimental data on an event-by-event basis, allowing one to track
all correlations in the five-dimensional phase space.
The event-based PWA of the $\gamma p\to \pi^0\eta p$ CBELSA/TAPS data by the Bonn-Gatchina (BnGa)
 group is a good example of such
 a technique~\cite{CBELSA_pi0etap_2008,CBELSA_pi0etap_2010,CBELSA_pi0etap_2014,BGPWA}.
Another method of analyzing a reaction like $\gamma N\to \pi^0\eta N$
is a simultaneous fit of various experimentally measured distributions for observables
sensitive to the reaction dynamics. A special model for the analysis of three-body final states,
especially aiming for understanding the features of $\gamma N\to \pi\pi N$ and $\gamma N\to \pi^0\eta N$,
was developed by the Mainz group~\cite{Mainz_2005,Mainz_2008,Mainz_2010,Mainz_2011,Mainz_2013},
paying particular attention to the analysis of specific angular distributions,
 which could be measured experimentally. The experimental data presented in this paper
 are compared to the previous solutions from the BnGa and Mainz groups, as well as
to a new fit with the revised Mainz model.

The $\gamma N\to \pi^0\eta N$ reaction recently became a subject of specific interest,
after a new analysis of the data acquired earlier with the GRAAL facility observed a narrow structure 
in the invariant-mass spectrum of the $\eta N$ system that could be interpreted as a contribution from a $N(1685)$ state~\cite{GRAAL_2017}.
The largest signal was observed in the $\gamma p\to \pi^0\eta p$ reaction, but the statistical accuracy of the measurement was quite low.
This mass range attracted much attention after the observation of a narrow bump
in the $\gamma n\to \eta n$ excitation function near the center-of-mass (c.m.) energy
$W=1.68$~GeV~\cite{GRAAL_etan_2007,CBELSA_etan_2008,A2_etan_2013,A2_etan_2014,A2_etan_2016},
while the $\gamma p\to \eta p$ total cross section showed a dip at the same
energy~\cite{etamamic,etaprmamic}. So far this effect, which was called a ``neutron anomaly'',
has no unique explanation in various partial-wave analyses (PWA)~\cite{BGPWA,WitthauerEtaN}.
Meanwhile, the latest analysis of the available $\gamma p\to \eta p$ data with
the revised $\eta$MAID model describes the dip at $W=1.68$~GeV well, without
 introducing any narrow state~\cite{etaprmamic}.

The most reliable identification of the $\gamma p\to \pi^0\eta p$ reaction in the A2 setup
comes by detecting its four-photon final state, with $\eta$ mesons decaying into two photons.
For the $\eta\to 3\pi^0$ decay mode, the experimental acceptance drops significantly and there
is a large chance of misidentifying the $\eta$ meson in the $4\pi^0$ final state.
Although the four-photon final state also has a large contribution from
the $\gamma p\to \pi^0\pi^0 p$ production, the kinematic-fit technique allows a reliable
 separation of the $\pi^0\eta\to 4\gamma$ final state from $\pi^0\pi^0\to 4\gamma$.
The previous analyses of the $\gamma p\to \pi^0\eta p$ A2
 data~\cite{Kashev_pi0eta_prod,Kashev_pi0eta_asym,Mainz_2010} were based on
the information available after the initial reconstruction of the detected events,
and all observables for $\gamma p\to \pi^0\eta p$ were measured by fitting
the $\eta\to 2\gamma$ signal above the background remaining mostly from
 $\gamma p\to \pi^0\pi^0 p$ events. Such an approach provides a poorer experimental
 resolution, compared to using the kinematic fit,
and does not allow making any PWA on the event-by-event basis.
Because the $\gamma p\to \pi^0\eta p$ reaction has three particles in its final state,
the event-by-event fit of the data is much more efficient than fitting separate
spectra measured for individual observables.

The A2 data used in the present analysis were taken in 2007 (Run I) and 2009 (Run II).
The production properties of the $\gamma p\to \pi^0\eta p$ reaction,
which are not related to polarization observables, were earlier
presented in Ref.~\cite{Kashev_pi0eta_prod}, based on the analysis of Run I.
The circular beam asymmetry for this reaction was reported for the first time
 in Ref.~\cite{Kashev_pi0eta_asym}, based on the analysis of Run II.
In the present work, all results are obtained from both Run I and Run II, after using
 a kinematic-fit technique for event identification and reconstruction.
The same technique was used previously in the analyses of the same data sets for measuring
 the reactions $\gamma p\to \eta p$~\cite{etamamic,etaprmamic},
$\gamma p\to \pi^0\pi^0 p$~\cite{Kashevarov2012}, $\gamma p\to K^0\Sigma^+$~\cite{K0Sigpl2013},
and $\gamma p\to \omega p$~\cite{a2_omegap_2015}.
 Compared to previous $\gamma p\to \pi^0\eta p$ measurements, the present analysis
 improves both the statistical accuracy, with a total of
 $1.5\times 10^6$ accumulated events, and the data quality, allowing
 a finer energy binning in the measurement of all observables needed for
 understanding the reaction dynamics. The data from Run II, which were taken with a higher beam energy, 
 provide the $\eta N$ invariant-mass distribution with good statistical accuracy in the vicinity of 1.685 GeV. 
 This makes it possible to search for a narrow structure, the observation of which was reported in Ref.~\cite{GRAAL_2017}.

\section{Experimental setup}
\label{sec:Setup}

The $\gamma p\to \pi^0\eta p$ reaction
was measured by using the Crystal Ball (CB)~\cite{CB}
as a central calorimeter and TAPS~\cite{TAPS,TAPS2}
as a forward calorimeter. These detectors were
installed in the energy-tagged bremsstrahlung photon beam of
the Mainz Microtron (MAMI)~\cite{MAMI,MAMIC}.
The photon energies were determined
by the Glasgow tagging spectrometer~\cite{TAGGER,TAGGER1,TAGGER2}.

The CB detector is a sphere consisting of 672
optically isolated NaI(Tl) crystals, shaped as
truncated triangular pyramids, which point toward
the center of the sphere. The crystals are arranged in two
hemispheres that cover 93\% of $4\pi$, sitting
outside a central spherical cavity with a radius of
25~cm, which contains the target and inner
detectors. In this experiment, TAPS was
arranged in a plane consisting of 384 BaF$_2$
counters of the hexagonal cross section.
It was installed 1.5~m downstream of the CB center
and covered the full azimuthal range for polar angles
from $1^\circ$ to $20^\circ$.
More details on the energy and angular resolution of the CB and TAPS
are given in Refs.~\cite{etamamic,slopemamic}.

 The present measurement used electron beams
 with energies of 1508 and 1557 MeV from the Mainz Microtron, MAMI-C~\cite{MAMIC}.
 The data with the 1508-MeV beam were taken in 2007 (Run I)
 and those with the 1557-MeV beam in 2009 (Run II).
 Bremsstrahlung photons, produced by the beam electrons
 in a 10-$\mu$m Cu radiator and collimated by a 4-mm-diameter Pb collimator,
 were incident on a liquid hydrogen (LH$_2$) target located
 in the center of the CB. The LH$_2$ target was 5-cm and 10-cm long
 in Run I and Run II, respectively.

 The target was surrounded by a particle identification
 (PID) detector~\cite{PID} used to distinguish between charged and
 neutral particles. It is made of 24 scintillator bars
 (50-cm long, 4-mm thick), arranged as a cylinder with a radius of 12 cm.

 The energies of the incident photons were analyzed
 up to 1402~MeV in Run I and up to 1448~MeV in Run II,
 by detecting the postbremsstrahlung electrons
 in the Glasgow tagged-photon spectrometer
 (Glasgow tagger)~\cite{TAGGER,TAGGER1,TAGGER2}.
 The uncertainty in the energy of the tagged photons is mostly determined
 by the segmentation of the tagger focal-plane detector in combination with
 the energy of the MAMI electron beam used in the experiments.
 Increasing the MAMI energy increases the energy range covered
 by the spectrometer and also has the corresponding effect on the uncertainty
 in $E_\gamma$. For both the MAMI energy settings of 1508 and 1557~MeV,
 this uncertainty was about $\pm 2$~MeV.
 More details on the tagger energy calibration and the corresponding uncertainties
 can be found in Ref.~\cite{TAGGER2}.

 The experimental trigger in Run I required the total energy deposited in the CB
 to exceed $\sim$320~MeV and the number of so-called hardware clusters
 in the CB (multiplicity trigger) to be two or more.
 In the trigger, a hardware cluster in the CB was a block of 16
 adjacent crystals in which at least one crystal had an energy
 deposit larger than 30 MeV.
 Depending on the data-taking period, events with a cluster multiplicity
 of two were prescaled with different rates.
 TAPS was not included in the multiplicity trigger for these experiments.
 In Run II, the trigger on the total energy
 in the CB was increased to $\sim$340~MeV, and the multiplicity
 trigger required $\ge 3$ hardware clusters in the CB.

\section{Data handling}
\label{sec:Data}

 The events from the $\gamma p\to \pi^0\eta p$ reaction were searched for in
 the four-photon final state produced via the $\eta\to 2\gamma$ decay mode.
 The $\gamma p\to 4\gamma p$ candidates were extracted from events with four or
 five clusters reconstructed in the CB and TAPS together by a software analysis.
 Five-cluster events were analyzed by assuming that all final-state particles
 had been detected, and four-cluster events by assuming that the detected particles 
 were photons.

 Because another strong contribution to the four-photon final state comes
 from the $\gamma p\to \pi^0\pi^0 p$ reaction, the latter events have to
 be separated from $\gamma p\to \pi^0\eta p\to 4\gamma p$ events.
 To identify these two processes, both hypotheses
 were tested with a kinematic fit and its output was used
 to reconstruct the reaction kinematics.
 Details of the kinematic-fit parametrization of the detector
 information and resolutions are given in Ref.~\cite{slopemamic}.
 The selection criteria for $\gamma p\to \pi^0\eta p\to 4\gamma p$ events
 were optimized by using Monte Carlo (MC) simulations of both
 reactions and from knowing the production rate of
 $\gamma p\to \pi^0\pi^0 p$~\cite{Kashevarov2012} with respect to the $\pi^0\eta$
 final state. The selection criteria were optimized to leave less
 than 1\% of the $\gamma p\to \pi^0\pi^0 p$ background remaining
 in the selected $\gamma p\to \pi^0\eta p\to 4\gamma p$ events
 within the entire energy range of the measurement.
 The analysis of the MC simulations showed that, for both the reactions that
 contribute to the four-photon final state, there are events for which
 both tested hypotheses give a reasonable probability.
 It was found that the contribution from $\gamma p\to \pi^0\pi^0 p$ becomes less
 than 1\% after applying the following criteria for the kinematic-fit probabilities:
 the probability $P$ for the $\gamma p\to \pi^0\eta p\to 4\gamma p$
 kinematic-fit hypothesis had to be larger than 3\% for five-cluster events and
 larger than 8\% for four-cluster events, while $P(\gamma p\to \pi^0\pi^0 p\to 4\gamma p)$
 had to be less than 0.1\% for both cluster multiplicities.

 The background remaining in the selected $\gamma p\to \pi^0\eta p\to 4\gamma p$ events
 originated from only two sources, which could both be directly subtracted from
 the experimental spectra. The first background was from interactions of
 the bremsstrahlung photons in the windows of the target cell.
 The subtraction of this background was based on the
 analysis of data samples that were taken with an empty
 (no liquid hydrogen) target cell. The statistical weight for the subtraction
 of the empty-target spectra was taken as a ratio of the photon-beam
 fluxes for the data samples with the full and the empty target.
 The second background was caused by random coincidences
 between tagger counts and experimental triggers.
 The subtraction of this background was carried out by using
 event samples for which all coincidences were random
 (see Refs.~\cite{etamamic,slopemamic} for more details).

 The MC simulations of the $\gamma p\to \pi^0\pi^0 p$ reaction were based
 on a previous study of this reaction with the same data sets~\cite{Kashevarov2012},
 which, in the given energy range, revealed the dominance of the
 $\gamma p \to \Delta(1232) \pi^0 \to \pi^0\pi^0 p$ process,
 with a smaller contribution from $\gamma p \to D_{13}(1520) \pi^0 \to \pi^0\pi^0 p$.
 The tests carried out for both processes showed the same efficiency for
 eliminating the $\gamma p\to \pi^0\pi^0 p$ reaction with the above selection cuts.

 The simulations of $\gamma p\to \pi^0\eta p\to 4\gamma p$ events used
 four models: phase space, $\gamma p \to \Delta(1232) \eta \to \pi^0\eta p$,
 $\gamma p \to S_{11}(1535)\pi^0 \to \pi^0\eta p$,
 and $\gamma p \to a_0(980) p \to \pi^0\eta p$, with the $\Delta(1232)$,
 $S_{11}(1535)$, and $a_0(980)$ Breit-Wigner (BW) parameters taken from
 the Review of Particle Physics (RPP)~\cite{PDG}.
 All production angles and the resonance decay distributions were generated isotropically.

 For all reactions, the generated events
 were propagated through a {\sc GEANT} (version 3.21) simulation of the experimental
 setup. To reproduce the resolutions observed in the experimental data, the {\sc GEANT}
 output (energy and timing) was subject to additional smearing,
 allowing both the simulated and experimental data to be analyzed in the same way.
 Matching the energy resolution between the experimental and MC events
 was achieved by adjusting the invariant-mass resolutions,
 the kinematic-fit stretch functions (or pulls), and probability
 distributions. Such an adjustment was based on the analysis of the
 same data sets for reactions having almost no background
 from other physical reactions
 (namely, $\gamma p\to \pi^0 p$, $\gamma p\to \eta p\to \gamma\gamma p$,
 and $\gamma p\to \eta p\to 3\pi^0p$~\cite{slopemamic}).
 The simulated events were also tested to check whether they passed
 the trigger requirements.

\section{Mainz model}
\label{sec:Model}
The theoretical analysis of the present $\gamma p\to \pi^0\eta p$ data was made within the framework of
 a revised isobar model, developed earlier by the Mainz group. In this model,
 the photoproduction amplitude consists of the three parts: (i) the Born amplitude,
 (ii) the resonant terms, and (iii) additional background contributions
\begin{equation}\label{12th}
t=t^B+t^R+t^{Bc}\,.
\end{equation}
The first two terms are basically similar to those used in the earlier model version and
 the analysis~\cite{Mainz_2010} of the data from Run I.
 The Born term $t^B$ contains the diagrams in which the intermediate state includes
a virtual nucleon or the $N(1535)1/2^-$ resonance in the direct or the crossed channel
(see diagrams $(a)$ to $(f)$ in Fig.~2 of Ref.~\cite{Mainz_2010}).

The resonance part $t^R$ is represented by the standard nonrelativistic BW form.
Based on the two dominant $\pi\eta N$ decay modes, the resonance amplitude can be represented as
the intermediate formation of the two quasi-two-body states $\eta\Delta(1232)$ and $\pi N(1535)$
\begin{equation}\label{15th}
t^R=t^{R(\eta\Delta)}+t^{R(\pi N^*)}\,.
\end{equation}
Each term in Eq.~(\ref{15th}) has the form
\begin{eqnarray}\label{20th}
t_{m_f\lambda}^{R(\alpha)}&=&\sum_{R(J^\pi)}
\frac{A^R_\lambda\,f_{R\alpha}}{W-M_R+\frac{i}{2}
\Gamma^R_{tot}(W)}\\
&\times& F^{J^\pi(\alpha)}\,
\Omega^{J^\pi}_{m_f\lambda}(\Omega_\pi,\Omega_\eta,\Omega_p)\,,\quad \alpha=\eta\Delta,\ \pi N^*\,,\nonumber
\end{eqnarray}
where the quantum numbers of the initial and final states are the total $\gamma N$ helicity $\lambda$
and the $z$ projection $m_f$ of the final-nucleon spin. The summation in Eq.\,(\ref{20th}) is over
the $\Delta$-like resonance states $R(J^\pi)$, determined by their spin-parity $J^\pi$ and
 having the BW masses and widths $M_R$ and $\Gamma^R_{tot}$.
$A_\lambda^R$ are the helicity functions of the transition $\gamma p\to R$ with $\lambda=1/2,\,3/2$.
 For the $\Delta(1700)3/2^-$ resonance, $A_\lambda^R$ is energy dependent according to
\begin{equation}\label{25th}
A^R_\lambda(W)=A^R_\lambda(M_R)\left(
\frac{\omega_\gamma}{\omega_\gamma^R}\right)^{3/2}\,,
\end{equation}
where $\omega_\gamma$ is the c.m. photon energy and $\omega_\gamma^R$ is its energy calculated at the resonance
position $W=M_R$.

The coupling constants $f_{R\alpha}$ in Eq.\,(\ref{20th}) determine the decay of the resonance $R$ into
 the quasi-two-body channel $\alpha=\eta\Delta,\,\pi N^*$. Depending on the invariant energies
 $\omega_{\eta N}$ and $\omega_{\pi N}$ of the $\eta N$ and $\pi N$ subsystems, the factors $F^{J^\pi(\alpha)}$
 are
\begin{eqnarray}\label{30th}
&&F^{J^\pi(\eta\Delta)}=\frac{f_{\Delta\pi
N}}{m_\pi^{L_\eta+1}}\ G_\Delta(\omega_{\pi N})\,q^{L_\eta}_\eta\, q_\pi^*,\\
&& \nonumber\\
\label{31th} &&F^{J^\pi(\pi N^*)}=\frac{f_{N^*\eta
N}}{m_\pi^{L_\pi}}\ G_{N^*}(\omega_{\eta N})\,q^{*\,L_\pi}_\pi,\ \quad\
\end{eqnarray}
 where $G_\Delta$ and $G_{N^*}$ are the $\Delta$ and $N^*$ propagators that have the same nonrelativistic BW form.
 The quantum numbers $L_\eta$ and $L_\pi$, determined by $J^\pi$, are the relative orbital angular momenta
 associated with the $\eta \Delta$ and $\pi N^*$ decays of the resonance $R$.
 The functions $\Omega^{J^\pi}_{m_f\lambda}$ in Eq.\,(\ref{20th}) describe the full angular dependence
 of the transition amplitude $t_{m_f\lambda}^{R(\alpha)}$.

 Compared to the previous model~\cite{Mainz_2010},
 the photoproduction amplitude from Eq.~(\ref{12th}) now includes a new term that represents
 the background amplitude $t^{Bc}$.
 This term involves only the lowest partial waves with $J\leq 5/2$ and, as described below, was treated in
 a phenomenological manner. The major constraint of the theory is that the $t^{Bc}$ contribution
 should be small, wherever possible, and should have a weak energy dependence.
 Although the background term does not have a simple physical picture, by introducing this term,
 it is accepted that there is no present theory capable of correctly predicting nonresonant
 contributions in the lower partial waves of the reaction under study.
 Indeed, in the earlier analysis of Ref.~\cite{Mainz_2010}, the nonresonant part of
 the photoproduction amplitude was represented only by the Born amplitude $t^B$,
 whereas, in the BnGa model~\cite{BGPWA}, the known dominant Regge exchanges were used instead.
 In addition, including the $t^{Bc}$ term effectively takes into account possible contributions from
 resonances with larger masses, which are not included in the model but can affect our energy range
 through their BW tails.

The parametrization of the $t^{Bc}$ term is similar to Eqs.\,(\ref{15th}) and (\ref{20th})
\begin{equation}\label{40th}
t^{Bc}=t^{Bc(\eta\Delta)}+t^{Bc(\pi N^*)}\,,
\end{equation}
but with the BW dependence replaced with energy-dependent functions $f_{J^\pi}^{(\alpha)}(W)$:
\begin{equation}\label{45th}
t_{m_f\lambda}^{Bc(\alpha)}=\sum_{J^\pi}
f_{J^\pi}^{(\alpha)}(W)\,F^{J^\pi(\alpha)}\Omega_{m_f\lambda}^{J^\pi}(\Omega_\pi,\Omega_\eta,\Omega_p)\,.
\end{equation}
To determine the energy dependence of the $t^{Bc}$ terms,
 the data were first fitted with only four principal resonances:
$\Delta(1700)3/2^-$, $\Delta(1905)5/2^+$, $\Delta(1920)3/2^+$, and $\Delta(1940)3/2^-$,
which, according to analyses reported in Refs.~\cite{Mainz_2010,BGPWA}, dominate in
the given energy range. The parameters of these four resonances were fixed to the results
obtained for them in Ref.~\cite{Mainz_2010}.
With such an input, the amplitudes $t^{Bc}$ were adjusted by fitting the data independently
for each individual energy bin. After the full energy range was covered,
the energy dependence of these background amplitudes was analyzed in each partial
wave $J^\pi$ to look for a resonance-like behavior.
Where the function $f_{J^\pi}^{(\alpha)}$ from Eq.~(\ref{45th}) demonstrated strong variation with energy,
an additional BW resonance was introduced into the amplitude $J^\pi$.
 The parameters of the new resonances were then determined during the subsequent fit to the data,
 performed over the full energy range. The parameters of the four principal resonances were also allowed to vary
 during the second fit. Free resonance parameters included $M_R$, $A_\lambda^R(M_R)$,
 $\sqrt{\beta_{\eta\Delta}}A_{1/2}$ , $\sqrt{\beta_{\pi N^*}}A_{1/2}$, and the ratio $A_{3/2}/A_{1/2}$,
 with quantities $\beta_\alpha=\Gamma^R_{(\alpha)}/\Gamma^R_{tot}$ for $\alpha=\eta\Delta,~\pi N^*$
 are the partial decay widths for $R\to \alpha$.
 As explained previously in Ref.~\cite{Mainz_2010}, because the resonance amplitudes depend on
 the product of the electromagnetic and hadronic vertices, the helicity functions $A^R_\lambda$
 and the partial decay widths $\Gamma^R_{(\alpha)}$ cannot be determined separately with reliable accuracy,
 forcing the use of their products $\sqrt{\beta_\alpha}A_{1/2}$ together with the ratio of $A_{3/2}/A_{1/2}$.
 The total widths of resonances $\Gamma_{tot}^R$ were fixed in the fits to their magnitudes from RPP~\cite{PDG}
 or previous PWAs~\cite{CBELSA_pi0etap_2008,CBELSA_pi0etap_2010,CBELSA_pi0etap_2014,BGPWA}.
 As explained in Ref.~\cite{Mainz_2010}, the reason for fixing these values lies in
 the closeness of the resonances, especially the $\Delta(1700)3/2^-$ state, to the reaction threshold,
 so that the experimental data do not provide sufficient constraints to extract resonance widths from the fit.

 To assure the smooth energy dependence of the background amplitudes remaining
 after introducing the new resonances, the functions $f_{J^\pi}^{(\alpha)}(W)$ in Eq.~(\ref{45th})
 were parametrized in terms of polynomials of order 2 (where the nonessential index $J^\pi$ is omitted)
\begin{equation}\label{55th}
f^{(\alpha)}(W)=\sum_{n=0}^2C_n\left(\frac{W}{M_N+m_\eta+m_\pi}\right)^n\,,
\end{equation}
with complex coefficients $C_n$.

 Speaking of the reliability of the present fit to the data,
 it is well known that, with limited polarization data, the fit solution may not be unique.
 Therefore, a rapid change in the background parameters within a narrow energy range could occur,
 not because of a real resonance, but owing to an accidental jump from one solution to another
 similar solution. Such a possibility was not investigated systematically here.
 However, the possible existence of alternative solutions was studied by varying
 the initial parameters for the four principal resonances mentioned above.
 In the end, although initial single-energy fits were
 yielding quite different background amplitudes, the subsequent fits, with all model parameters released,
 converged to the same solution.

\section{Results and discussion}
\label{sec:Results}
\begin{figure*}
\includegraphics[width=0.98\textwidth]{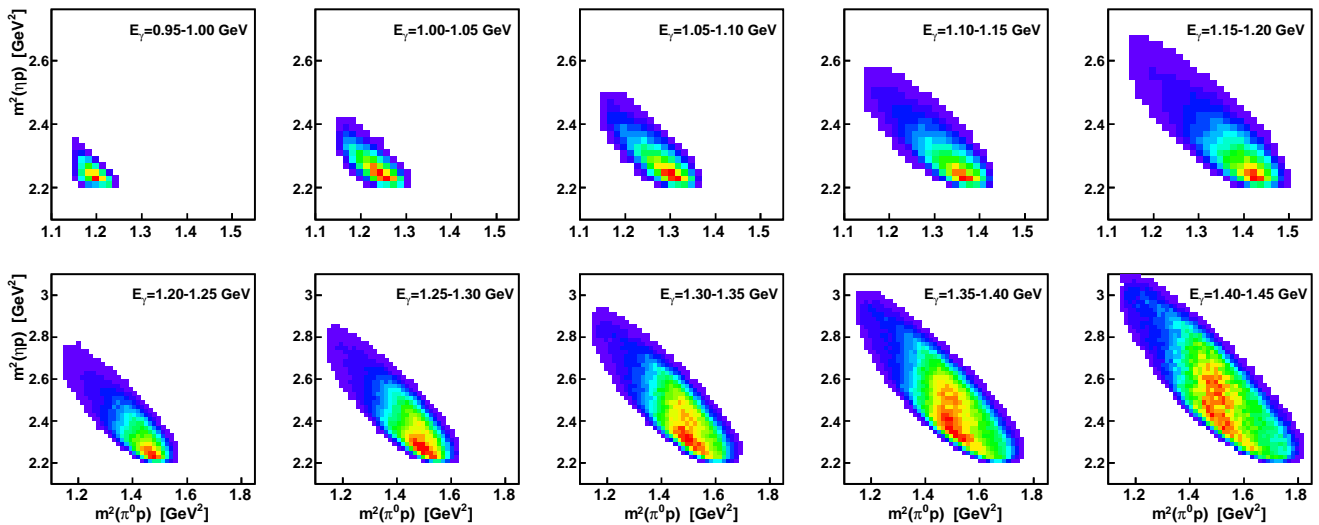}
\caption{
 Experimental Dalitz plots of $m^2(\eta p)$ vs $m^2(\pi^0 p)$ combined from Run I and Run II,
 shown for 10 energy intervals between $E_{\gamma}=0.95$~GeV and $1.45$~GeV.
}
 \label{fig:dalpl_pi0etap_a2_exp}
\end{figure*}
\begin{figure*}
\includegraphics[width=0.98\textwidth]{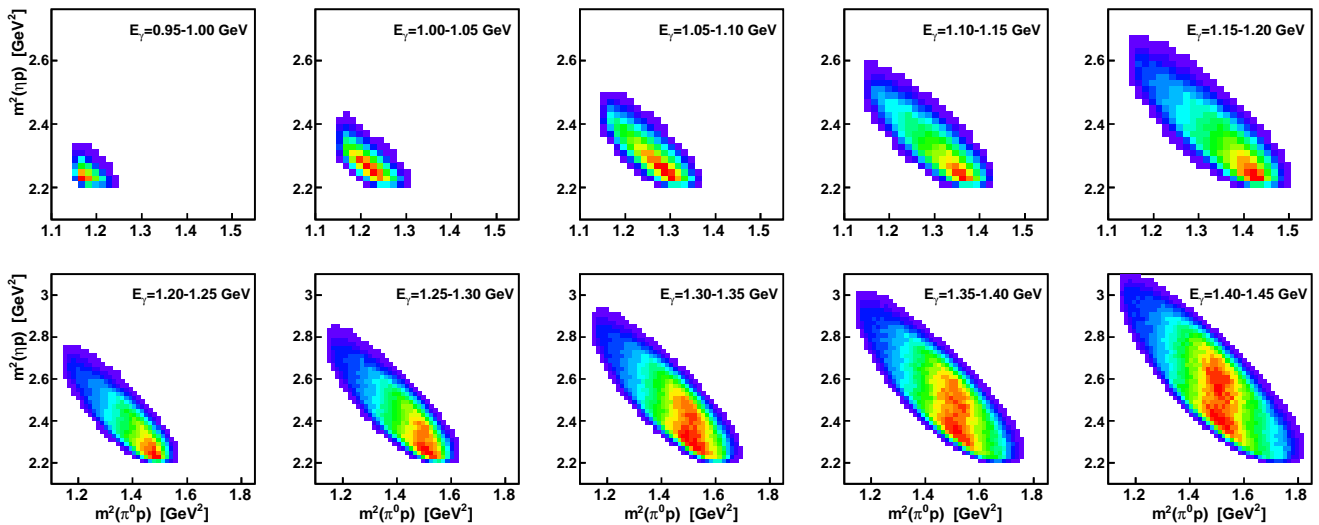}
\caption{
 Same as Fig.~\protect\ref{fig:dalpl_pi0etap_a2_exp},
 but for MC simulation of $\gamma p \to \Delta(1232) \eta \to \pi^0\eta p$.
}
 \label{fig:dalpl_pi0etap_mc_delta_apr09}
\end{figure*}

\begin{figure*}
\includegraphics[width=0.98\textwidth]{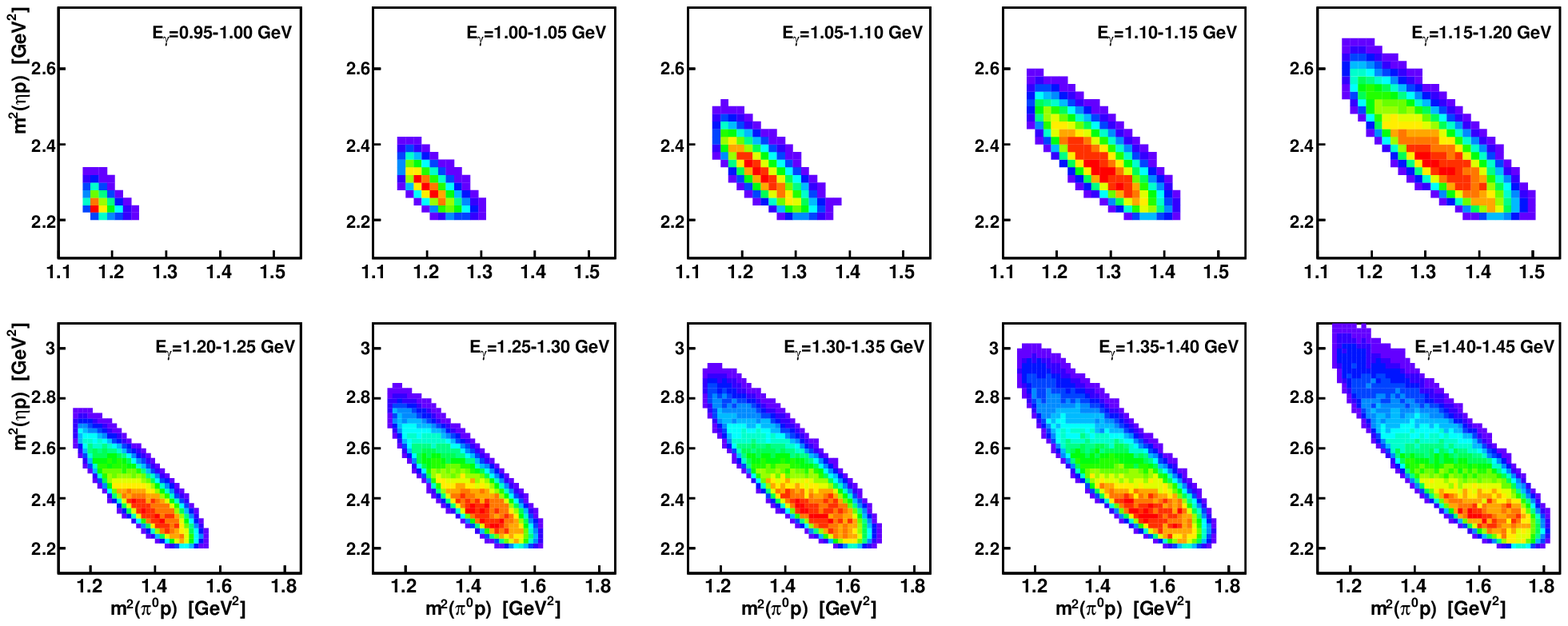}
\caption{
 Same as Fig.~\protect\ref{fig:dalpl_pi0etap_a2_exp},
 but for MC simulation of $\gamma p \to N(1535)\pi^0 \to \pi^0\eta p$.
}
 \label{fig:dalpl_pi0etap_mc_s11_apr09}
\end{figure*}

\begin{figure*}
\includegraphics[width=0.98\textwidth]{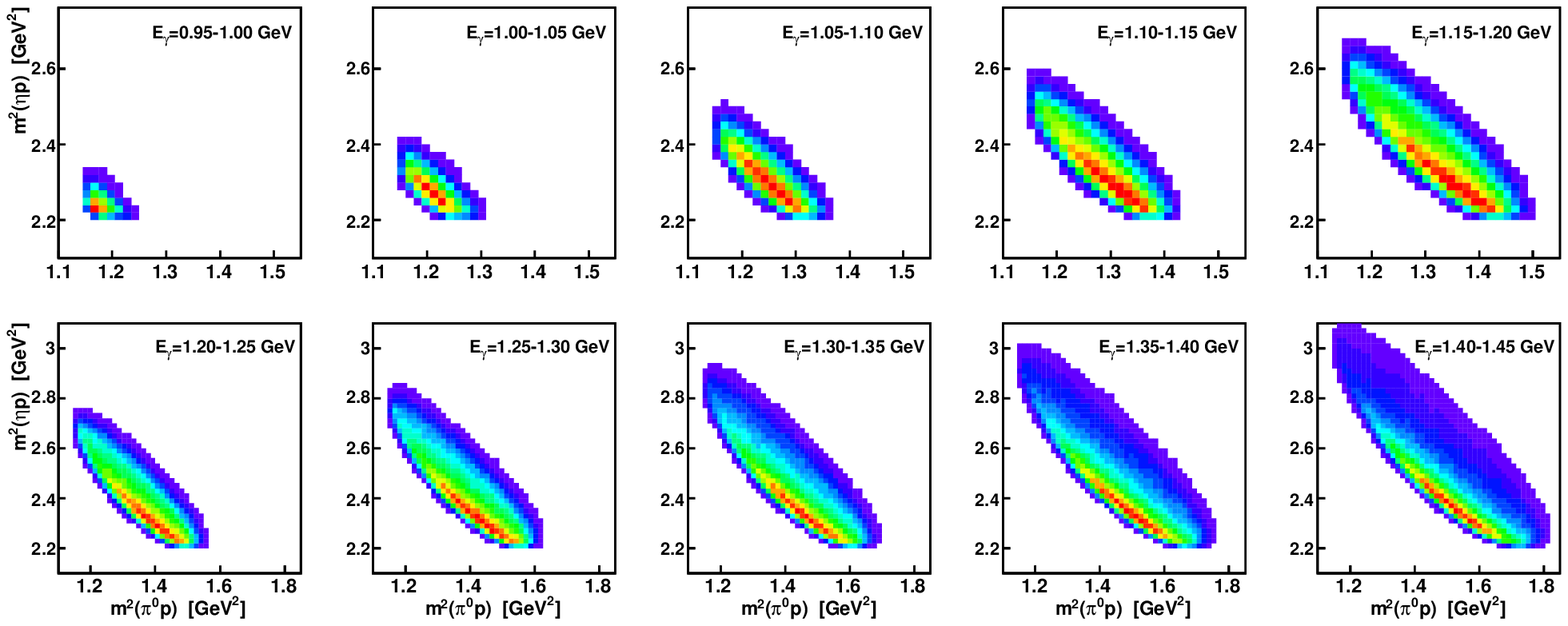}
\caption{
 Same as Fig.~\protect\ref{fig:dalpl_pi0etap_a2_exp},
 but for MC simulation of $\gamma p \to a_0(980)p \to \pi^0\eta p$.
}
 \label{fig:dalpl_pi0etap_mc_a0_apr09}
\end{figure*}

\begin{figure*}
\includegraphics[width=0.98\textwidth]{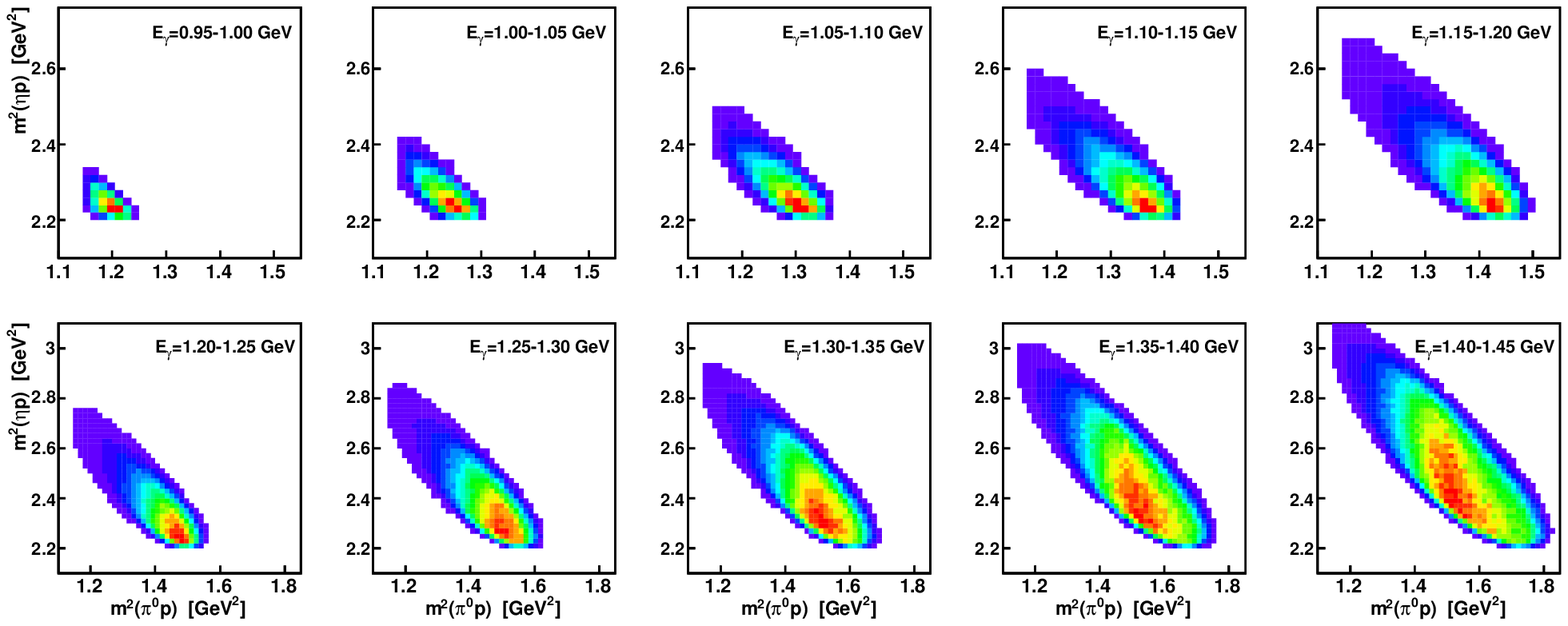}
\caption{
 Same as Fig.~\protect\ref{fig:dalpl_pi0etap_a2_exp},
 but for the phase-space MC simulation weighted with BnGa PWA
 of the CBELSA/TAPS data~\protect\cite{CBELSA_pi0etap_2014}.
}
 \label{fig:dalpl_pi0etap_mc_bnga_apr09}
\end{figure*}

 One of the most informative distributions for the three-body final-state reactions
 are their Dalitz plots and the energy dependence of their density distributions.
 In Fig.~\ref{fig:dalpl_pi0etap_a2_exp}, the experimental Dalitz plots of
 $m^2(\eta p)$ vs $m^2(\pi^0 p)$, obtained by combining the results of Run I and Run II together,
 are shown for ten 50-MeV-wide incident-photon energy intervals in the range
 from $E_{\gamma}=0.95$~GeV to $1.45$~GeV.
 In Figs.~\ref{fig:dalpl_pi0etap_mc_delta_apr09},~\ref{fig:dalpl_pi0etap_mc_s11_apr09},
 and~\ref{fig:dalpl_pi0etap_mc_a0_apr09}
 the corresponding Dalitz plots are shown, respectively, for the MC simulations of
 $\gamma p \to \Delta(1232) \eta \to \pi^0\eta p$, $\gamma p \to N(1535)\pi^0 \to \pi^0\eta p$,
 and $\gamma p \to a_0(980) p \to \pi^0\eta p$, where all resonance decays were generated isotropically.
 As seen from the comparison with Fig.~\ref{fig:dalpl_pi0etap_a2_exp},
 the production of the $\pi^0\eta p$ final state in the given energy range
 occurs mostly via the $\Delta(1232) \eta$ intermediate state. The difference in the density along
 the $\Delta$ band reflects the deviation from the isotropic distribution of $\Delta$
 decay products, used in the MC simulation, with respect to the $\Delta$ direction.
 The higher density of the experimental $\Delta$ band at low  $m(\eta p)$ masses
 means that the $\pi^0$ mesons from $\Delta \to \pi^0 p$ decays are produced more
 in the direction of the $\Delta$.
 The contribution from $N(1535)1/2^-$ seems to be significantly smaller than from $\Delta(1232)$.
 According to the analysis of CBELSA/TAPS data~\cite{CBELSA_pi0etap_2014},
 the contribution from $\gamma p \to a_0(980)p\to \pi^0\eta p$ reaches a few percent
 at energies around $E_{\gamma}=1.4$~GeV. However, as seen from the Dalitz plots,
 the visual observation of such a contribution is complicated because the $a_0(980)$ band
 overlaps with the $\Delta(1232)$ band at this energy.
 The phase-space MC simulation weighted with BnGa PWA of the CBELSA/TAPS
 data~\cite{CBELSA_pi0etap_2014}, illustrated in Fig.~\ref{fig:dalpl_pi0etap_mc_bnga_apr09},
 demonstrates reasonable agreement with the experimental plots of the present work.
\begin{figure*}
\includegraphics[width=0.98\textwidth]{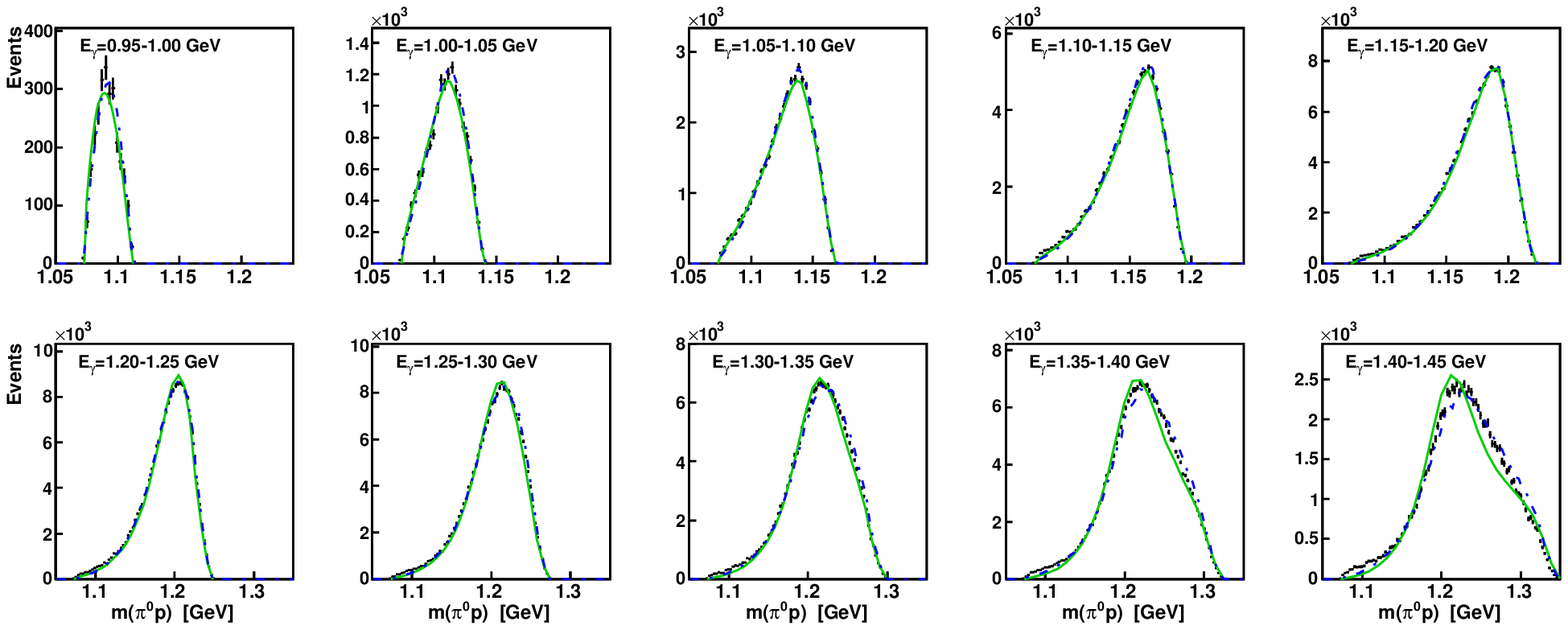}
\caption{
 Comparison of the invariant-mass $m(\pi^0p)$ spectra combined from
 Run I and Run II (crosses) with
 BnGa PWA~\protect\cite{CBELSA_pi0etap_2014} (blue dash-dotted line)
 and the Mainz model used to fit the present data (solid green line).
}
 \label{fig:mpi0p_pi0etap_a2_exp_th}
\end{figure*}
 For a better comparison with theoretical analyses,
 the present results for the invariant-mass spectra of $m(\pi^0p)$, $m(\eta p)$,
 and $m(\pi^0\eta)$ are compared in Figs.~\ref{fig:mpi0p_pi0etap_a2_exp_th},
 \ref{fig:metap_pi0etap_a2_exp_th}, and~\ref{fig:mpi0eta_pi0etap_a2_exp_th}
 with the BnGa PWA of CBELSA/TAPS data~\cite{CBELSA_pi0etap_2014}
 and with the Mainz model used to fit the present data.
 As seen, the agreement of the experimental results with both BnGa PWA and Mainz model
 is quite good, with some discrepancies appearing only for highest energies.
 The earlier solution of the Mainz model~\cite{Kashev_pi0eta_prod,Mainz_2010} is not shown here, as
 the invariant-mass distributions were not included in that fit.

 In addition, it is important to note that, the Dalitz plots and invariant mass spectra at the highest photon energy range
 shown in Figs.~\ref{fig:dalpl_pi0etap_a2_exp} and~\ref{fig:metap_pi0etap_a2_exp_th} do not clearly indicate any narrow structure in the vicinity of 
 $m(\eta p) = 1.685$~GeV (or $m^{2}(\eta p)=2.84$~GeV$^{2}$) reported in Ref.~\cite{GRAAL_2017}. However, it is worth noting that, 
 compared to the present measurement, the data in Ref.~\cite{GRAAL_2017} cover higher photon-energy range, reaching $E_\gamma =1.5$~GeV.
 Further search for a potential narrow state in the region of $m(\eta p) = 1.685$~GeV or determination of the corresponding upper limit 
 is beyond the topic of the present paper.

\begin{figure*}
\includegraphics[width=0.98\textwidth]{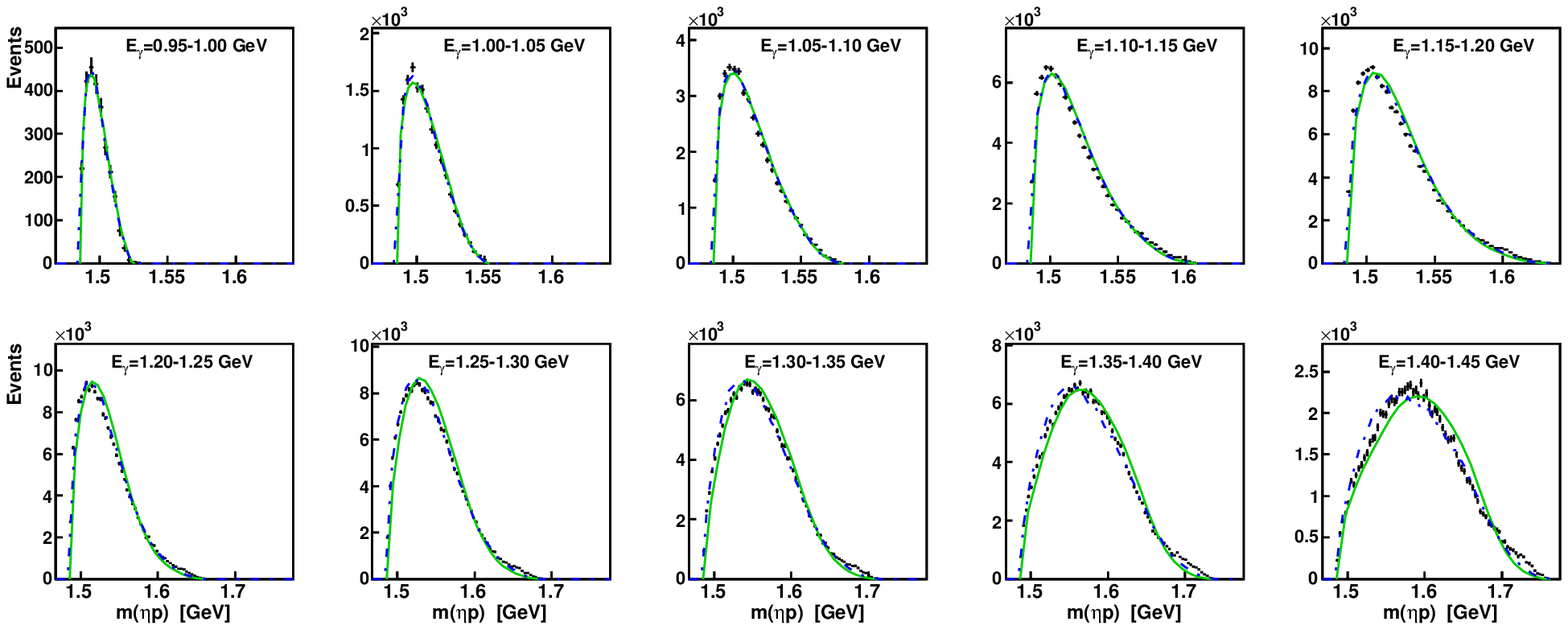}
\caption{
  Same as Fig.~\protect\ref{fig:mpi0p_pi0etap_a2_exp_th},
 but for the invariant mass $m(\eta p)$.
}
 \label{fig:metap_pi0etap_a2_exp_th}
\end{figure*}
\begin{figure*}
\includegraphics[width=0.98\textwidth]{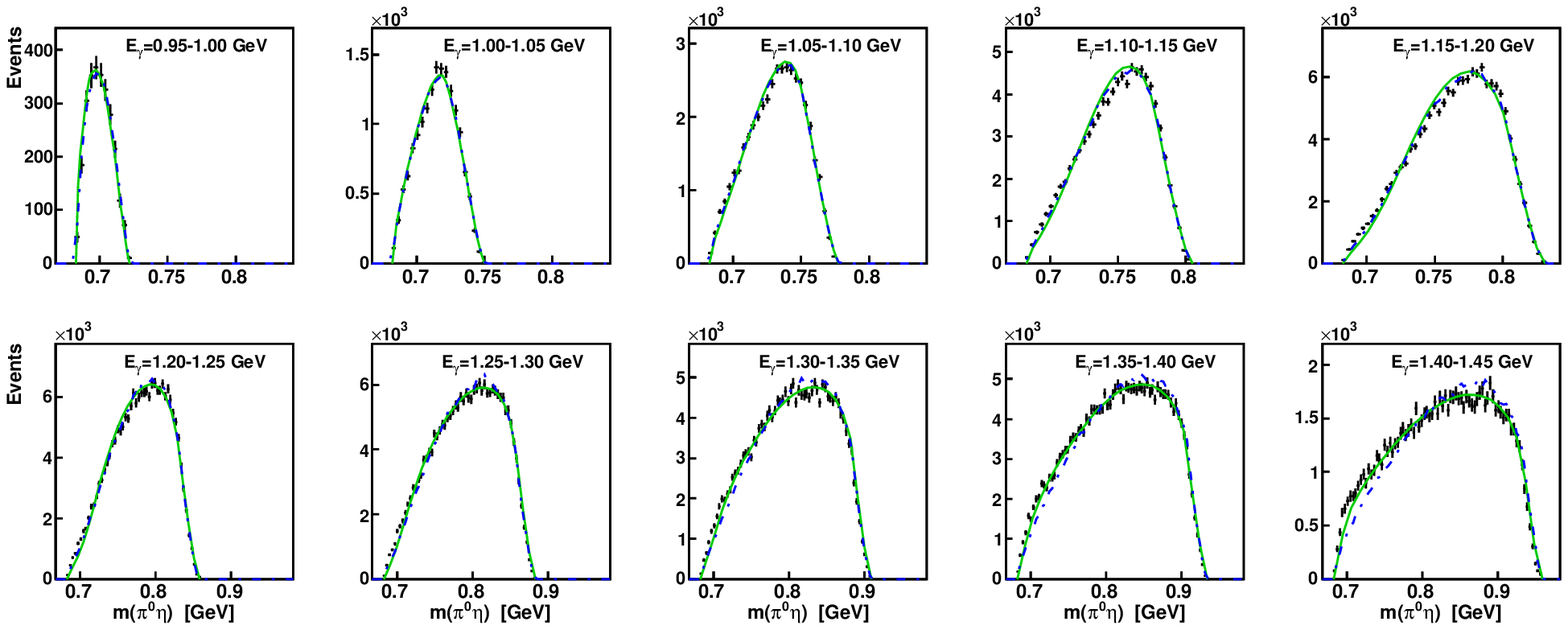}
\caption{
  Same as Fig.~\protect\ref{fig:mpi0p_pi0etap_a2_exp_th},
 but for the invariant mass $m(\pi^0\eta)$.
}
 \label{fig:mpi0eta_pi0etap_a2_exp_th}
\end{figure*}

 Because of the five-dimensional phase space of the $\gamma p \to \pi^0\eta p$ reaction,
 guided by the results for the experimental Dalitz plots and invariant-mass spectra,
 the acceptance determination was based on MC simulations for
 $\gamma p \to \Delta(1232) \eta \to \pi^0\eta p$,
 with a small fraction of phase space added. Thus, the following results for the total
 and differential cross sections include some systematic uncertainties
 stemming from approximations in the acceptance correction.
 Based on the tests with different MC simulations,
 such systematic uncertainties were estimated to be around 5\% at the lowest energies,
 increasing to 8\% at the largest energies.
 A more precise determination of the reaction dynamics
 could be made with a PWA based on the event-by-event data, and this work will provide
 such data for future PWAs.

\begin{figure*}
\includegraphics[width=0.98\textwidth]{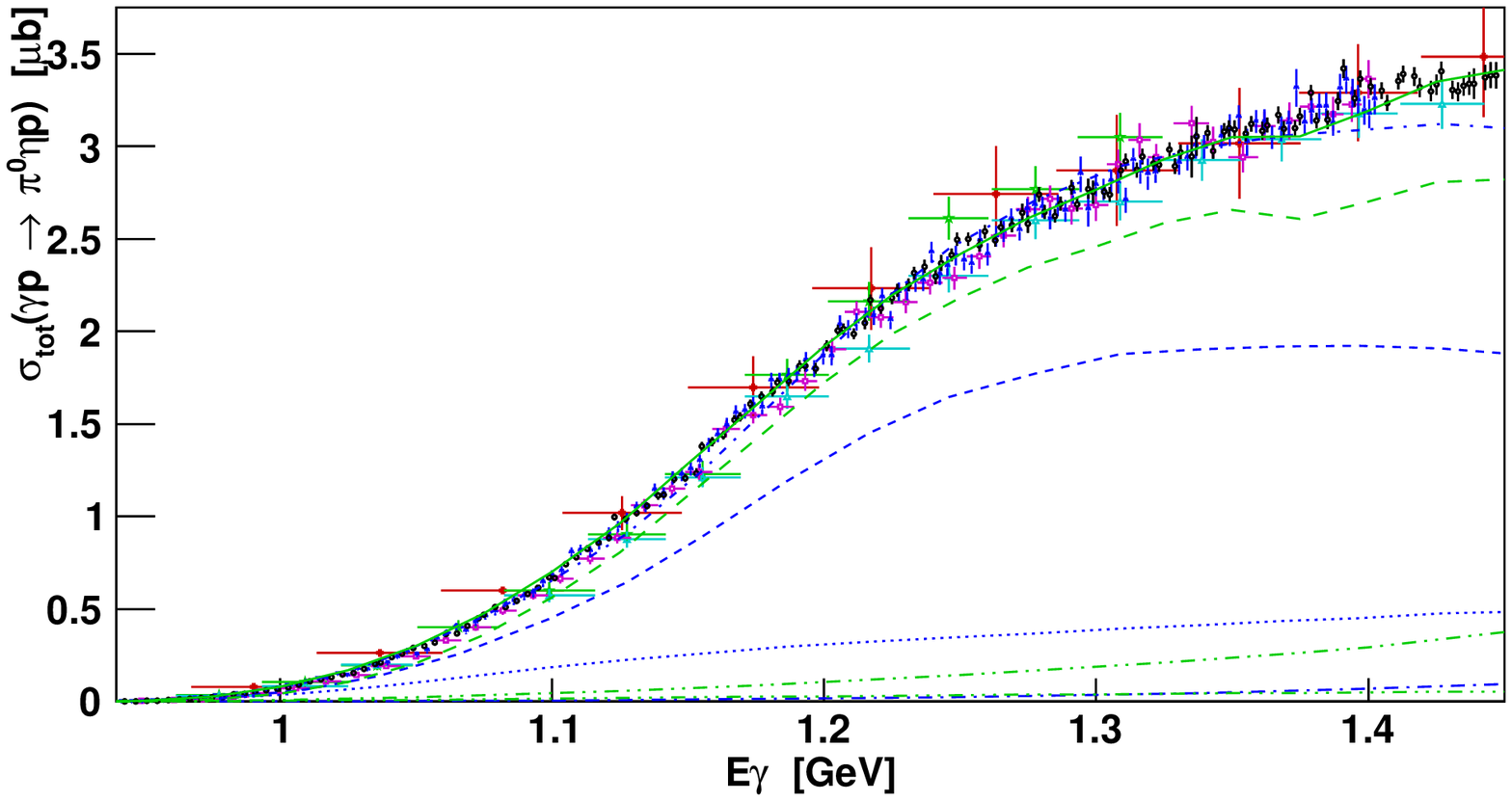}
\caption{
 Comparison of the $\gamma p\to \pi^0\eta p$ total cross sections from Run I (blue solid triangles)
 and Run II (black open circles) with each other and with previous data by
 CBELSA/TAPS~\protect\cite{CBELSA_pi0etap_2008,CBELSA_pi0etap_2014}
 (green stars and cyan open triangles, respectively),
 GRAAL~\protect\cite{GRAAL_2008} (red crosses),
 and A2~\protect\cite{Kashev_pi0eta_prod} (magenta open squares).
 The predictions from BnGa PWA~\protect\cite{CBELSA_pi0etap_2014} are shown by blue lines
 for the total cross section (dash-dotted) and individual contributions from
 $\gamma p \to \Delta(1232) \eta$ (dashed), $\gamma p \to N(1535)\pi^0$ (dotted),
 and $\gamma p \to a_0(980) p$ (long-dash-dotted).
 The results from analysis with the Mainz model are shown by green lines
 for the total cross section (solid) and individual contributions from
 the resonant (long-dashed), background (dash-double-dotted), and Born (dash-triple-dotted) terms.
}
 \label{fig:tcs_etapi0_a2_exp_th}
\end{figure*}

 The $\gamma p\to \pi^0\eta p$ total cross sections from Run I and Run II,
 obtained for the energies of each tagger channel above the reaction threshold,
 are shown in Fig.~\ref{fig:tcs_etapi0_a2_exp_th}, illustrating good agreement
 with each other as well as with previous data and theoretical analyses.
 In addition to the total cross section, the individual contributions
 from $\gamma p \to \Delta(1232) \eta$, $\gamma p \to N(1535)\pi^0$,
 and $\gamma p \to a_0(980) p$ are plotted for BnGa PWA~\cite{CBELSA_pi0etap_2014},
 confirming the dominance of the $\Delta(1232) \eta$ intermediate state,
 seen in the features of the experimental Dalitz plots.
 For the Mainz model, the individual contributions are plotted for
 Born ($t^B$), resonant ($t^R$), and background ($t^{Bc}$) terms from Eq.~(\ref{12th}),
 demonstrating the strong dominance of the resonant term.

 In the context of the Mainz model~\cite{Kashev_pi0eta_prod,Mainz_2010},
 the most informative observables for understanding the internal
 dynamics of the $\gamma p\to \pi^0\eta p$ production are angular distributions in
 any two-particle rest frame of the three-particle final state.
 Because the production is dominated by $\gamma p \to \Delta(1232) \eta$,
 the $\pi^0 p$ rest frame was chosen for measuring $\pi^0$ angular distributions
 in the canonical and helicity coordinate systems ($x',y',z'$).
 In the helicity system, the $z'$ axis was taken along the $\pi^0 p$ total momentum,
 the $y'$ axis was directed along the vector product of the $\eta$ and beam-photon vectors
 taken in the $\pi^0 p$ rest frame. The $x'$ axis is just a vector product of the $y'$ and $z'$ axes.
 In the canonical system, the $z'$ axis was taken along the beam-photon momentum in the c.m. frame,
 the $y'$ axis was directed along the vector product of the $\eta$ and beam-photon vectors
 also taken in the c.m. frame. In the CBELSA/TAPS analysis~\cite{CBELSA_pi0etap_2014}, the
 Gottfried-Jackson (GJ) frame was used instead of the canonical. In the GJ frame, all vectors
 used in the canonical system are taken in the $\pi^0 p$ rest frame, which makes the
 $\pi^0$ angular distributions very similar, but not identical, for these two systems.

 Similarly to the previous analysis of Run I~\cite{Kashev_pi0eta_prod,Mainz_2010},
 the distributions of $\cos\theta$ and angle $\phi$ of the $\pi^0$ meson were measured
 in both the helicity and canonical frames, where $\theta$ and $\phi$ are, respectively,
 the polar and azimuthal angles of the $\pi^0$ vector within the ($x',y',z'$)
 coordinate system. The present results include both
 Runs I and II data (doubling the statistics) and are divided into 10 energy bins,
 compared to the previous four.
 In the following comparison of the present results to previous measurements and
 predictions of different analyses, the $\cos\theta$ results by
 CBELSA/TAPS~\cite{CBELSA_pi0etap_2014} in the GJ frame are used to compare to
 the canonical-frame results ($\phi$ distributions were not published by CBELSA/TAPS).
 For a better comparison of the angular dependences, all the differential cross sections
 have been normalized so that their integrals equal one.

\begin{figure*}
\includegraphics[width=0.98\textwidth]{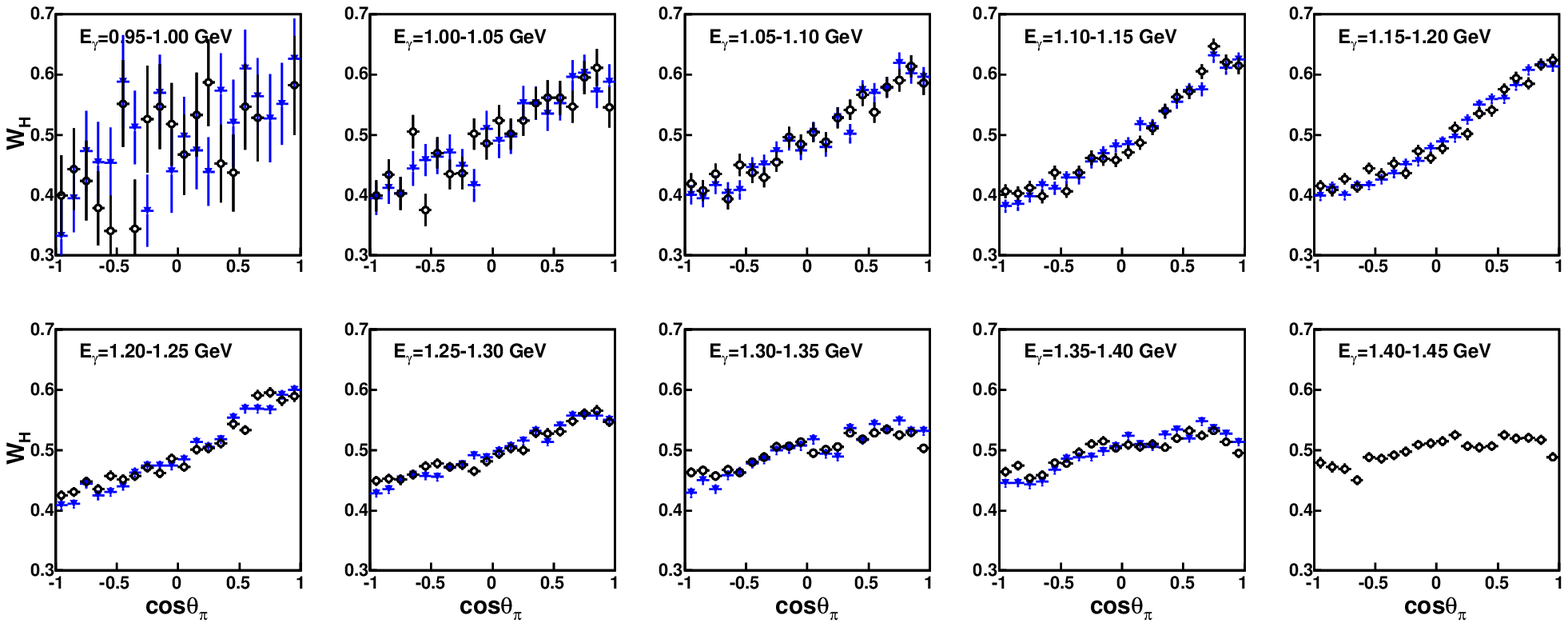}
\caption{
 Comparison of the results from Run I (blue triangles) and Run II (open circles)
 for the $\cos(\theta_{\pi^0})$ distributions in the helicity frame.
}
 \label{fig:costh_pi0_h_a2_2007_vs_09_pi0eta}
\end{figure*}

\begin{figure*}
\includegraphics[width=0.98\textwidth]{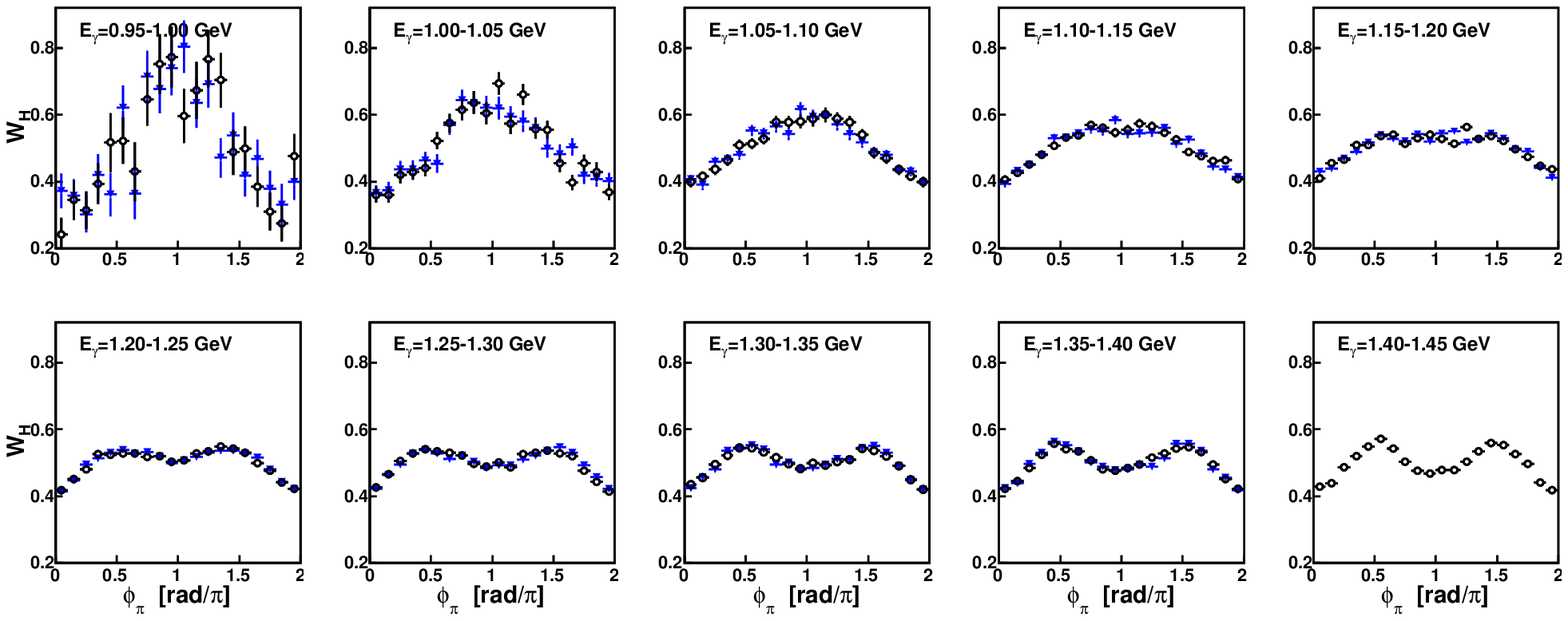}
\caption{
 Same as Fig.~\protect\ref{fig:costh_pi0_h_a2_2007_vs_09_pi0eta}, but for $\phi_{\pi^0}$.
}
 \label{fig:phi_pi0_h_a2_2007_vs_09_pi0eta}
\end{figure*}

 Before comparing the present angular distributions to various analyses,
 the consistency of the results obtained from Run I and Run II is illustrated in
 Figs.~\ref{fig:costh_pi0_h_a2_2007_vs_09_pi0eta} and \ref{fig:phi_pi0_h_a2_2007_vs_09_pi0eta} for
 the $\cos\theta$ and $\phi$ distributions obtained for the recoil $\pi^0$ in the helicity frame.
 As seen, the data points from both data sets are in good agreement within their statistical
 uncertainties. Because the highest-energy bin was not covered in Run I, the combined results
 are provided only for nine energies.

\begin{figure*}
\includegraphics[width=0.98\textwidth]{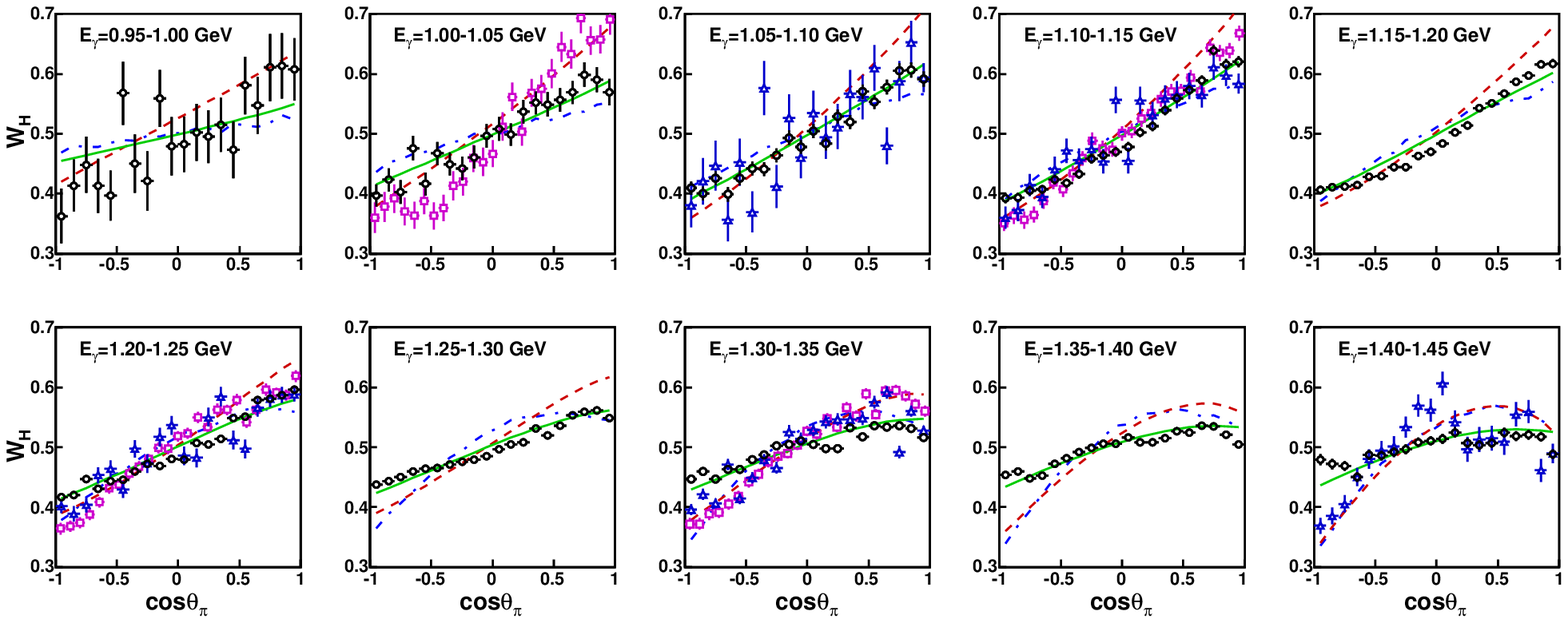}
\caption{
 Comparison of the present helicity-frame $\cos(\theta_{\pi^0})$ distributions,
 combined from Run I and Run II (open circles),
 to previous data by CBELSA/TAPS~\protect\cite{CBELSA_pi0etap_2008,CBELSA_pi0etap_2014}
 (blue stars, data points combined from both the references) and by
 A2~\protect\cite{Kashev_pi0eta_prod} (magenta open squares) at similar energies,
 and to predictions by BnGa PWA~\protect\cite{CBELSA_pi0etap_2014} (blue dash-dotted line)
 and by the earlier Mainz model~\cite{Mainz_2010} (red dashed line),
 and to the fit of the revised Mainz model to the present data (solid green line).
}
 \label{fig:costh_pi0_h_a2_vs_exp_th_pi0eta}
\end{figure*}

\begin{figure*}
\includegraphics[width=0.98\textwidth]{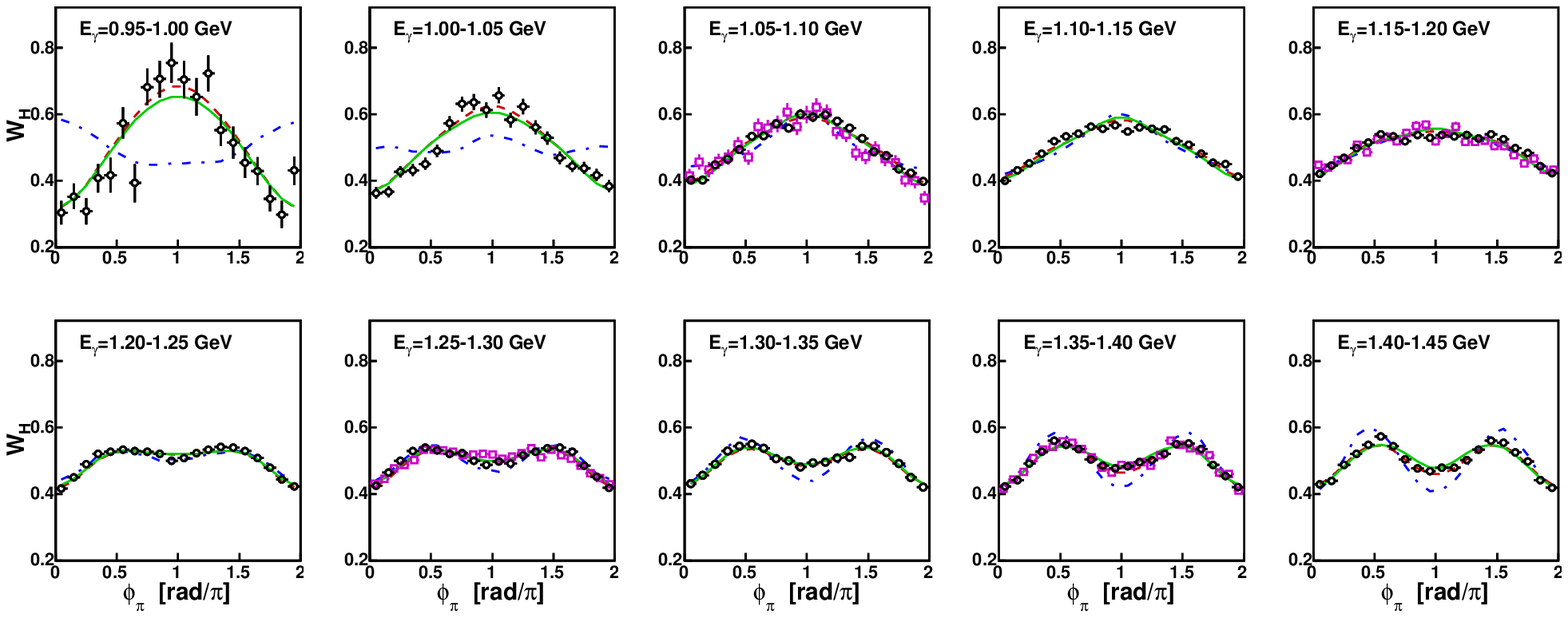}
\caption{
 Same as Fig.~\protect\ref{fig:costh_pi0_h_a2_vs_exp_th_pi0eta}, but for $\phi_{\pi^0}$.
}
 \label{fig:phi_pi0_h_a2_vs_exp_th_pi0eta}
\end{figure*}

 In Figs.~\ref{fig:costh_pi0_h_a2_vs_exp_th_pi0eta}
 and~\ref{fig:phi_pi0_h_a2_vs_exp_th_pi0eta}, the combined results
 for $\cos\theta$ and $\phi$ of the recoil $\pi^0$ in the helicity frame
 are compared to previous data at similar energies,
 to predictions by BnGa PWA~\cite{CBELSA_pi0etap_2014} and by the earlier Mainz model~\cite{Mainz_2010},
 and to the fit of the revised Mainz model to the present data.
 For better statistical accuracy, the CBELSA/TAPS data points from
 Refs.~\cite{CBELSA_pi0etap_2008,CBELSA_pi0etap_2014} have been combined together.
 As seen in Fig.~\ref{fig:costh_pi0_h_a2_vs_exp_th_pi0eta}, the present $\cos\theta$ distributions
 in the helicity frame are in better agreement with previous measurements and model analyses
 for lower energies, and the consistency with the CBELSA/TAPS data and
 BnGa PWA~\cite{CBELSA_pi0etap_2014} is better than with the previous
 A2 analysis~\cite{Kashev_pi0eta_prod,Mainz_2010}. The discrepancies are larger for
 the highest energies, where the results are more sensitive to the acceptance correction,
 which depends on the reaction dynamics used in the corresponding MC simulation.
 On the other hand, the revised Mainz model is able to describe
 the present data over the entire energy range.
 Compared to $\cos\theta$, the present $\phi$ distributions in the helicity frame are in
 much better agreement with the previous measurements and analyses,
 except BnGa PWA~\cite{CBELSA_pi0etap_2014} at the lowest energies,
 where the CBELSA/TAPS data have very low statistics.

\begin{figure*}
\includegraphics[width=0.98\textwidth]{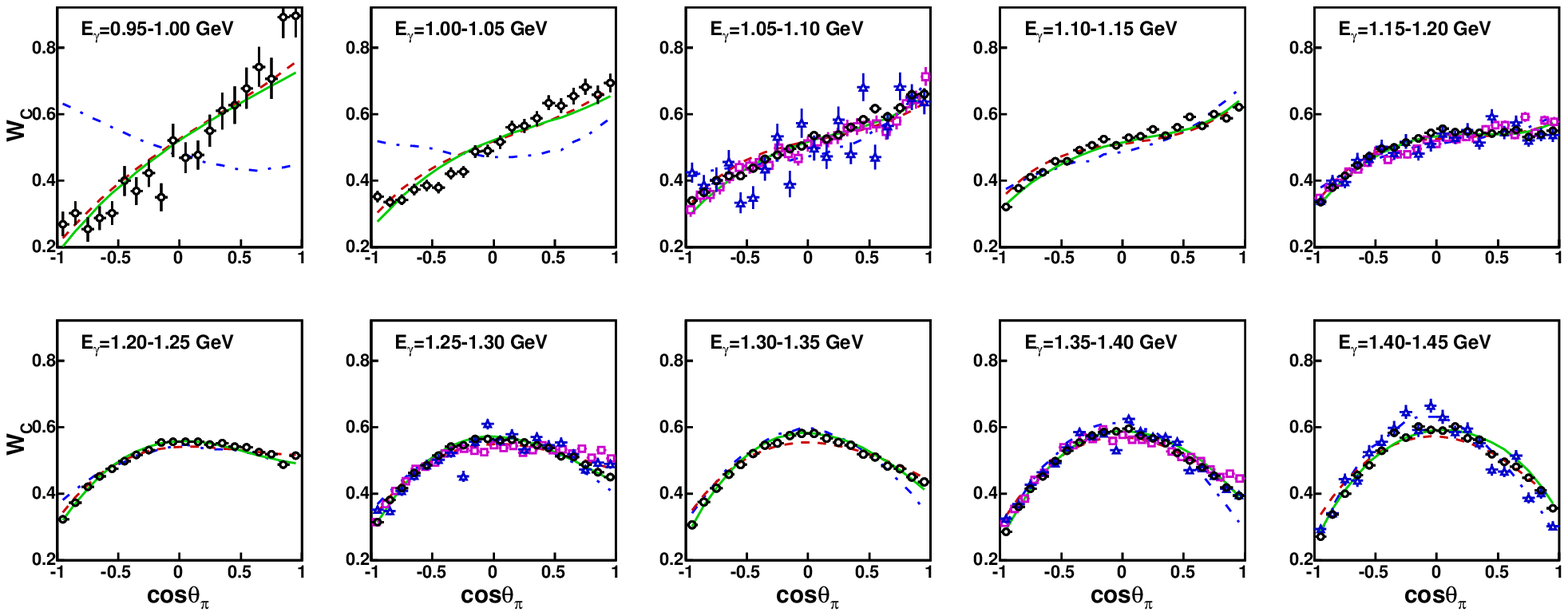}
\caption{
 Same as Fig.~\protect\ref{fig:costh_pi0_h_a2_vs_exp_th_pi0eta}, but for the canonical frame.
 The combined data from CBELSA/TAPS~\protect\cite{CBELSA_pi0etap_2008,CBELSA_pi0etap_2014}
 are shown for the GJ frame.
}
 \label{fig:costh_pi0_c_a2_vs_exp_th_pi0eta}
\end{figure*}

\begin{figure*}
\includegraphics[width=0.98\textwidth]{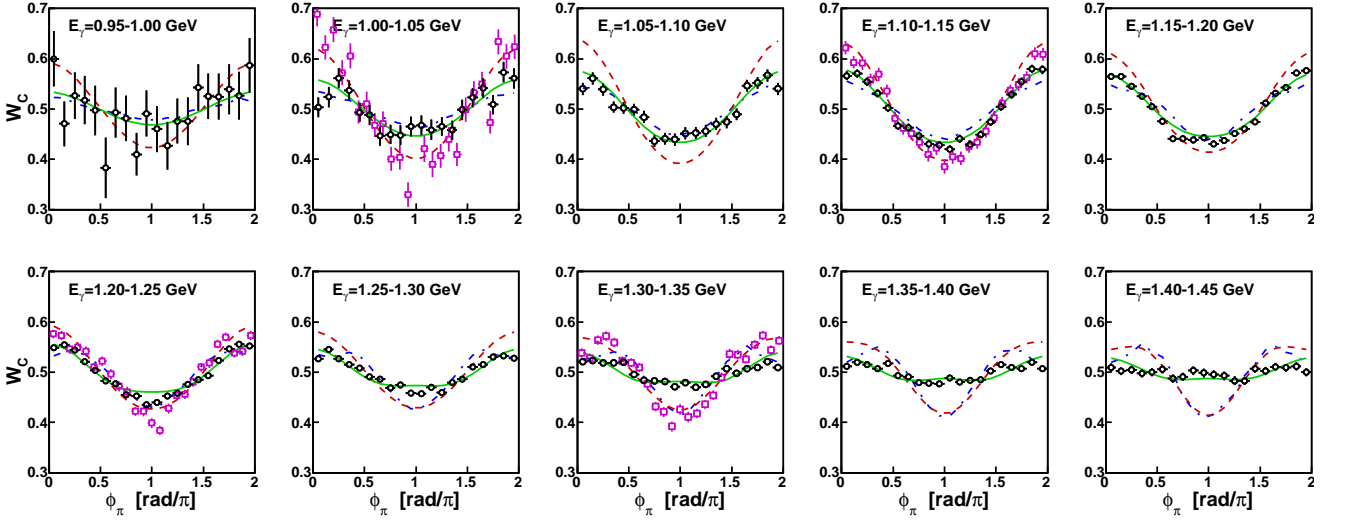}
\caption{
 Same as Fig.~\protect\ref{fig:costh_pi0_h_a2_vs_exp_th_pi0eta}, but for $\phi_{\pi^0}$
 in the canonical frame.
}
 \label{fig:phi_pi0_c_a2_vs_exp_th_pi0eta}
\end{figure*}

\begin{figure*}
\includegraphics[width=0.98\textwidth]{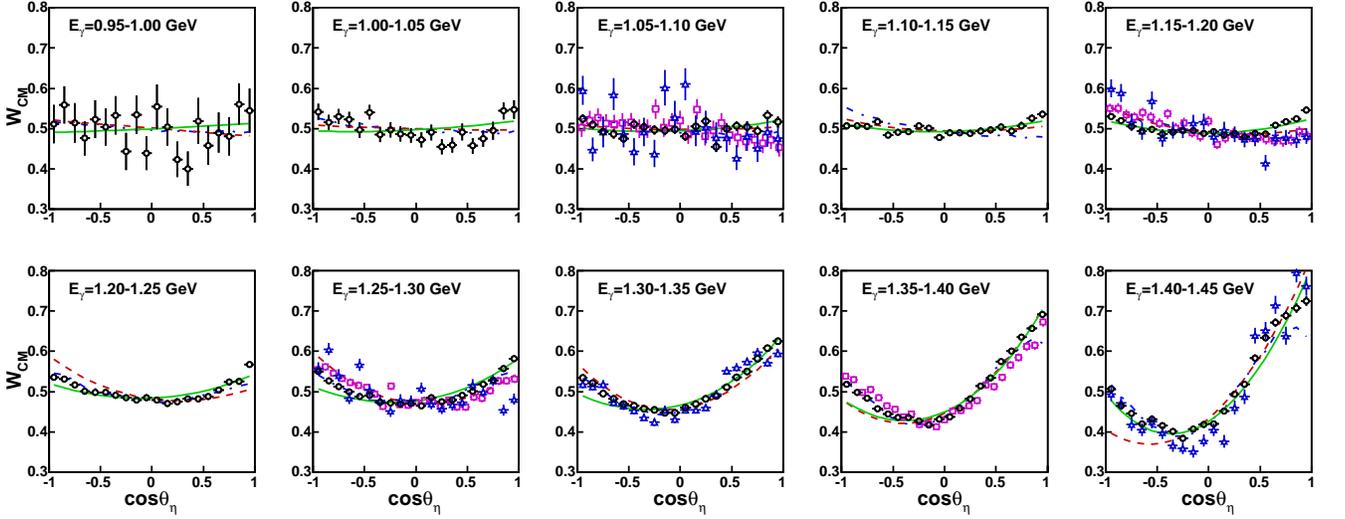}
\caption{
 Same as Fig.~\protect\ref{fig:costh_pi0_h_a2_vs_exp_th_pi0eta}, but for
 $\cos(\theta_\eta)$ spectra in the c.m. frame.
}
 \label{fig:costh_eta_cm_a2_vs_exp_th_pi0eta}
\end{figure*}

\begin{figure*}
\includegraphics[width=0.98\textwidth]{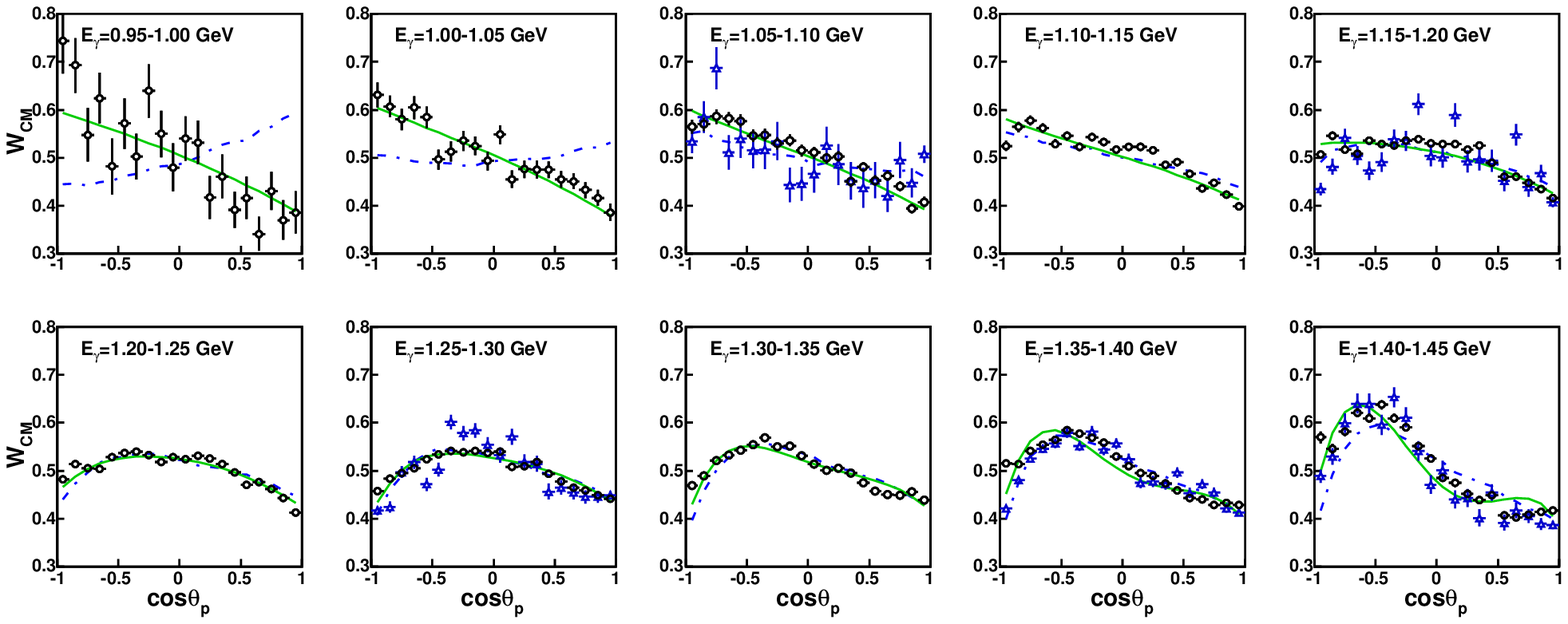}
\caption{
 Same as Fig.~\protect\ref{fig:costh_pi0_h_a2_vs_exp_th_pi0eta}, but for the recoil proton
 $\cos(\theta_p)$ spectra in the c.m. frame.
}
 \label{fig:costh_pr_cm_a2_vs_exp_th_pi0eta}
\end{figure*}

 The comparison of the present results in the canonical frame is given
 in Figs.~\ref{fig:costh_pi0_c_a2_vs_exp_th_pi0eta}
 and~\ref{fig:phi_pi0_c_a2_vs_exp_th_pi0eta} for $\cos\theta$ and $\phi$ of the recoil $\pi^0$,
 respectively. As seen, in contrast to the helicity frame, better agreement with previous
 measurements and analyses is obtained for $\cos\theta$, except BnGa PWA~\cite{CBELSA_pi0etap_2014}
 at the lowest energies. The present $\phi$ distributions in the canonical frame are in better agreement
 with previous measurements and model analyses for lower energies, and the consistency with
 BnGa PWA~\cite{CBELSA_pi0etap_2014} is better than with the previous
 A2 analysis~\cite{Kashev_pi0eta_prod,Mainz_2010}. The discrepancies seen in the highest energies
 could be caused by a stronger sensitivity of the results to the model used in the MC simulation
 to determine the experimental acceptance.
 The revised Mainz model describes the present data in the canonical frame for the entire energy range.
 The measurement of production angles of the final-state particles is presented for the $\eta$ and 
 proton, the c.m. $\cos\theta$ distributions of which are shown in Figs.~\ref{fig:costh_eta_cm_a2_vs_exp_th_pi0eta}
 and~\ref{fig:costh_pr_cm_a2_vs_exp_th_pi0eta}, respectively
 
 The corresponding distributions for $\pi^0$ are not shown as they are very similar to
 $\cos(\theta_{\pi^0})$ in the canonical or GJ frames.
 As seen, the present results for $\eta$ are in good agreement with previous measurements
 over almost the entire energy range, whereas the proton results contradict
 the predictions of the BnGa PWA~\cite{CBELSA_pi0etap_2014} near the reaction threshold.
 The $\cos\theta$ distributions for the recoil proton are not shown for the earlier analysis
 of the A2 data~\cite{Kashev_pi0eta_prod} as they were not extracted there.

\begin{figure*}
\includegraphics[width=0.98\textwidth]{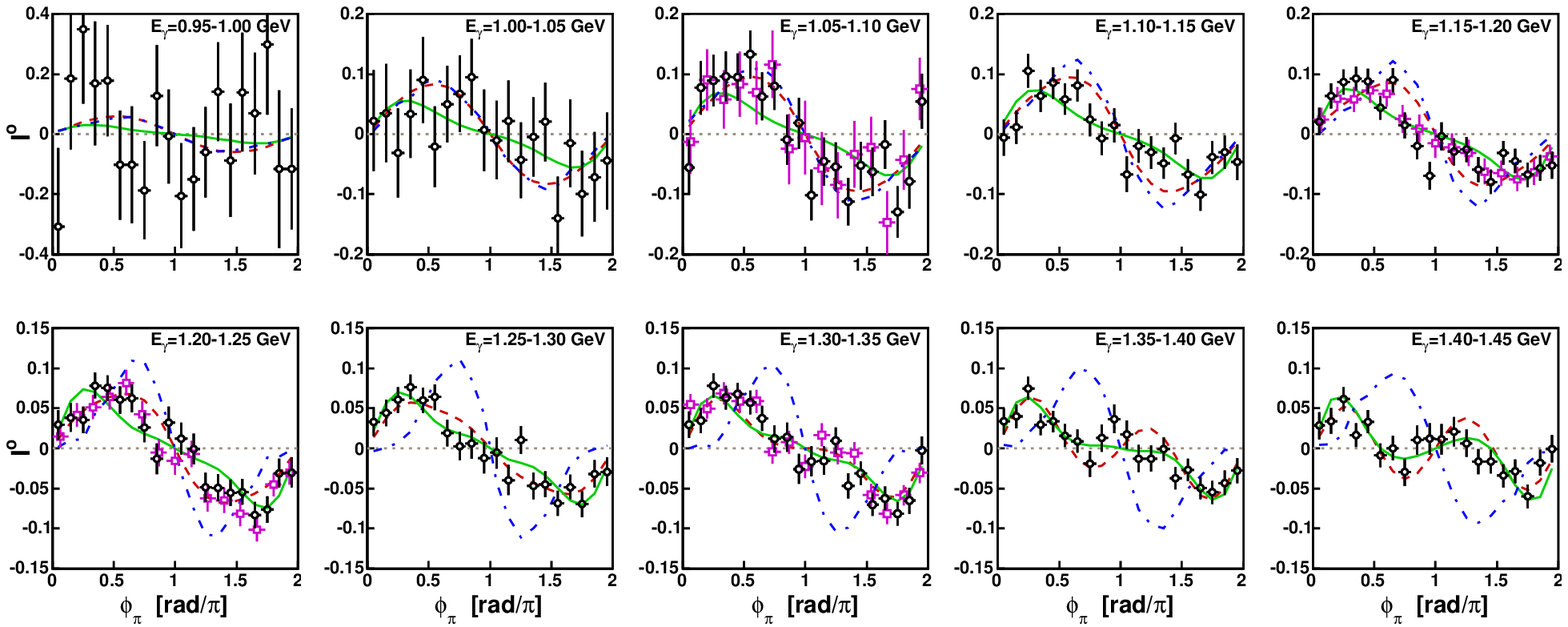}
\caption{
 Comparison of the present results for helicity photon asymmetry $I^\odot$ (open circles)
 to the previous analysis from Ref.~\protect\cite{Kashev_pi0eta_asym}
 (magenta open squares) at similar energies,
 to predictions by BnGa PWA~\protect\cite{CBELSA_pi0etap_2014} (blue dash-dotted line)
 and by the earlier Mainz model~\cite{Mainz_2010} (red dashed line),
 and to the fit with its revised version to the present data (solid green line).
}
 \label{fig:Io_asym_pi0eta_a2_vs_exp_th}
\end{figure*}

 In this work, the measurement of helicity photon asymmetry $I^\odot$ was made
 for 10 energy bins (the same as for the other observables), compared to four energy bins
 in Ref.~\cite{Kashev_pi0eta_asym}, the analysis of which was based on Run II only.
 In Fig.~\ref{fig:Io_asym_pi0eta_a2_vs_exp_th}, the present $I^\odot$ results
 are compared to the previous data from Ref.~\cite{Kashev_pi0eta_asym},
 to predictions by BnGa PWA~\cite{CBELSA_pi0etap_2014}, to
 the earlier Mainz model~\cite{Mainz_2010}, and to the fit with the revised model.
 As seen, the present results for $I^\odot$ are in good agreement with the previous
 data~\cite{Kashev_pi0eta_asym} within the error bars, whereas the fit with
 the revised Mainz model deviates from the earlier version.
 The discrepancy with the BnGa PWA~\cite{CBELSA_pi0etap_2014} is larger, and
 increases with energy.

 In summary, the present $\gamma p\to \pi^0\eta p$ data demonstrate better statistical accuracy,
 with finer energy binning, compared to previous measurements.
 The consistency of the present results with the earlier data and analyses is partial
 for some observables and energy ranges. For the most part, the observed discrepancies could
 be explained by the sensitivity of results to the five-dimensional acceptance correction and
 by poorer statistics and wider energy binning of the previous measurements.
 The discrepancies with the BnGa PWA~\cite{CBELSA_pi0etap_2014} are expected to be reduced
 by adding the present data into their new fit on the event-by-event basis.
 Such an analysis is now in progress and will be published by the BnGa group separately.
 Compared to the earlier Mainz model~\cite{Mainz_2010}, its revised version includes more
 observables in the fit and, as demonstrated in the figures, is able to describe
 their shape and energy dependences over the entire energy range.

 As discussed above, the earlier Mainz model~\cite{Mainz_2010} included
 only the first two terms of the amplitude~(\ref{12th}), used in the revised version.
 The main reason for introducing the purely phenomenological term $t^{Bc}$ was the fact
 that refitting the parameters of the earlier model to the entire set of the new results
 was not sufficient for a good description. After introducing the background amplitudes,
 it was found that the set of the four principal isobars
 ($\Delta(1700)3/2^-$, $\Delta(1905)5/2^+$, $\Delta(1920)3/2^+$, and $\Delta(1940)3/2^-$) 
 is sufficient for the resonance term $t^R$ to describe the data,
 and the $\Delta(1600)3/2^+$ and $\Delta(1750)1/2^+$ states, the contributions of which
 were less important in the analysis of Ref.~\cite{Mainz_2010}, were found to be unnecessary now.
 Also, similarly to the previous analysis~\cite{Mainz_2010}, there was no clear need to include
 resonances in the $1/2^-$ and $5/2^-$ waves to improve the data description.
 Though the contribution from the background term $t^{Bc}$ is considerably smaller than
 the resonant term $t^R$ (see Fig.~\ref{fig:tcs_etapi0_a2_exp_th}), its introduction
 improves the fit's $\chi^2/{\rm ndf}$ from 7.3 to 3.7, using the statistical uncertainties only.
 Another observation made from the fit to the present data is that the background amplitudes tend
 to cancel the Born amplitudes at higher energies, especially in the dominant $3/2^-$ wave.

 The resonance parameters obtained in the fit to the present data are listed
 in Table~\ref{tab}, along with the corresponding resonances and parameters obtained for them
 in Ref.~\cite{Mainz_2010}.  As seen, the parameters of the dominant resonance $\Delta(1700)3/2^-$
 are practically the same, whereas those of the other resonances changed.
 Similarly to Ref.~\cite{Mainz_2010}, the systematic uncertainties were not used in the fit with
 the revised Mainz model.
\begin{table*}
\caption{Parameters found for the resonances included in the revised Mainz model, compared
 in the second row with the corresponding values obtained in the earlier analysis of
 Ref.~\protect\cite{Mainz_2010}. The quantities $\beta_\alpha=\Gamma^R_{(\alpha)}/\Gamma^R_{tot}$
 for $\alpha=\eta\Delta,~\pi N^*$ are the partial decay widths for $R\to \alpha$.
}
\label{tab}
\begin{ruledtabular}
\begin{tabular}{|c|c|c|c|c|c|}
\hline
 $J^{\pi}[L_{2T2J}(M_R)]$ & $M_R$ & $\Gamma^R_{tot}(M_R)$
& $\sqrt{\beta_{\eta\Delta}}A_{1/2}(M_R)$ & $\sqrt{\beta_{\pi N^*}}A_{1/2}(M_R)$ & $A_{3/2}(M_R)/A_{1/2}(M_R)$   \\
  & [MeV] & [MeV] & [10$^{-3}$GeV$^{-1/2}$] & [10$^{-3}$GeV$^{-1/2}$] &  \\
\hline\hline
$\Delta(1700)3/2^-$  &  $1704\pm1$ & $375$ & $12.0\pm0.2$ & $8.4\pm 0.1$ & $0.80\pm 0.02$ \\
                     &  $1701\pm1$ & $375$ & $10.6\pm0.2$ & $8.9\pm 0.4$ & $0.95\pm 0.01$ \\
\hline
$\Delta(1905)5/2^+$  &  $1990\pm4$ & $330$ & $-44.8\pm0.5$ & $-1.5\pm 0.2$ & $-0.71\pm 0.02$ \\
                    &  $1873\pm4$ & $330$ & $-25.5\pm0.6$ & $-2.4\pm 0.4$ & $-0.70\pm 0.03$ \\
\hline
$\Delta(1920)3/2^+$  &  $1948\pm5$ & $260$ & $5.7\pm0.3$ & $2.1\pm 0.1$ & $4.40\pm 0.05$ \\
                     &  $1894\pm3$ & $200$ & $11.9\pm0.4$ & $4.4\pm 0.4$ & $1.15\pm 0.06$ \\
\hline
$\Delta(1940)3/2^-$  &  $1819\pm1$ & $450$ & $10.7\pm1.0$ & $3.5\pm 0.2$ & $2.30\pm 0.02$ \\
                     &  $1870\pm1$ & $450$ & $19.9\pm1.2$ & $9.3\pm 0.7$ & $1.65\pm 0.02$ \\
\hline
\end{tabular}
\end{ruledtabular}
\end{table*}

\section{Summary and conclusions}
\label{sec:Conclusion}

 The data available from the A2 Collaboration at MAMI were analyzed to select
 the $\gamma p\to \pi^0\eta p$ reaction on an event-by-event basis, which allows
 for partial-wave analyses of three-body final states to obtain more reliable results,
 compared to fits to measured distributions.
 These data provide the world’s best statistical accuracy in the energy range from
 threshold to $E_{\gamma}=1.45$~GeV, allowing a finer energy binning in the measurement
 of all observables needed for understanding the reaction dynamics.
 In this work, the $\gamma p\to \pi^0\eta p$ data are compared to the existing BnGA PWA and
 to the earlier Mainz model. The potential impact of the present data on future analyses was
 demonstrated by fitting these results with the revised Mainz model, which was able
 to describe all the differential cross sections and their energy dependences over
 the entire energy range. The invariant-mass distributions and Dalitz plots measured in this work for energies $E_\gamma <1.45$~GeV do not show
any clear indication for a narrow structure in the region of $m(\eta p) = 1.685$~GeV reported in Ref.~\cite{GRAAL_2017}.

\section*{Acknowledgment}

 The authors wish to acknowledge the excellent support of the accelerator group and
 operators of MAMI.
 We also thank A. Sarantsev and V. Nikonov on behalf of the BnGa PWA group.
 This work was supported by the Deutsche Forschungsgemeinschaft (SFB443,
 SFB/TR16, and SFB1044), DFG-RFBR (Grant No. 09-02-91330), the European Community-Research
 Infrastructure Activity under the FP6 ``Structuring the European Research Area''
 program (Hadron Physics, Contract No. RII3-CT-2004-506078), Schweizerischer
 Nationalfonds (Contracts No. 200020-156983, No. 132799, No. 121781, No. 117601, and No. 113511),
 the U.K. Science and Technology Facilities Council (STFC 57071/1, 50727/1),
the U.S. Department of Energy (Offices of Science and Nuclear Physics,
 Grants No. DE-FG02-99-ER41110, No. DE-FG02-88ER40415, and No. DE-FG02-01-ER41194),
 National Science Foundation (Grants No. PHY-1039130 and No. IIA-1358175),
 INFN (Italy), and NSERC (Canada, Grant No. FRN-SAPPJ-2015-00023). We also acknowledge the support of the Carl-Zeiss-Stiftung.
 A.~F. acknowledges additional support from the Tomsk Polytechnic University competitiveness enhancement program.
 We thank the undergraduate students of Mount Allison University
 and The George Washington University for their assistance.

\end{document}